\pdfoutput=1
\documentclass{article}
\usepackage{arxiv}

\usepackage[utf8]{inputenc} % allow utf-8 input
\usepackage[T1]{fontenc}    % use 8-bit T1 fonts
\usepackage{hyperref}       % hyperlinks
\usepackage{url}            % simple URL typesetting
\usepackage{booktabs}       % professional-quality tables
\usepackage{amsfonts}       % blackboard math symbols
\usepackage{nicefrac}       % compact symbols for 1/2, etc.
\usepackage{microtype}      % microtypography
\usepackage{lipsum}
\usepackage{graphicx}
\usepackage{moreverb,url}

\newcommand\BibTeX{{\rmfamily B\kern-.05em \textsc{i\kern-.025em b}\kern-.08em
T\kern-.1667em\lower.7ex\hbox{E}\kern-.125emX}}

\title{From form to information: 
\\Analysing built environments in different spatial cultures}

\author{Vinicius M. Netto \\
        Associate Professor \\
        Department of Urbanism\\
        Universidade Federal Fluminense (UFF)\\
        Visiting Scholar \\
        Center for Urban Science and Progress (CUSP NYU)\\
        \texttt{vmnetto@id.uff.br} \\
        \And 
        Edgardo Brigatti \\
        Associate Professor \\
        Institute of Physics\\
        Universidade Federal do Rio de Janeiro (UFRJ) \\
        \texttt{edgardo@if.ufrj.br} \\
        \And 
        Caio Cacholas \\
        Researcher \\
        Graduate Programme of Architectural and Urban Studies (PPGAU UFF)\\ }

\begin{document}

\maketitle

\begin{abstract}
Cities are different around the world, 
but does this fact have any relation to culture?
The idea that urban form embodies idiosyncrasies related to cultural identities captures the imagination of many in urban studies, but it is an assumption yet to be carefully examined. 
Approaching spatial configurations
in the built environment as a proxy of urban culture, this paper searches for differences potentially consistent with specific regional cultures or cultures of planning in urban development.
It does so focusing on the elementary components shaping cities: buildings and how they are aggregated in cellular complexes of built form.  
Exploring Shannon's work, we introduce an %new 
%\textcolor{green}{
entropy measure to analyse the probability distribution of cellular arrangements in built form systems.
We apply it to downtown areas of 45 cities from different regions of the world as a similarity measure to compare and cluster cities  
potentially consistent with specific spatial cultures.
Findings suggest a classification scheme that sheds further light on what we call the ‘cultural hypothesis’: the possibility that different cultures and regions find different ways of ordering space.
\end{abstract}


\section{Introduction}
\begin{flushright}
Cities of different cultural types and different scales embody different spatial identities... human societies order their spatial milieu in order to construct a spatial culture, that is, a distinctive way of ordering space.\\
Hillier \cite[p.5-6]{hillier1989architecture}
\end{flushright}

The idea that urban form embodies idiosyncrasies that express cultural identities seems to be a frequent assumption in urban studies. 
It has to do with the contextual role of custom and institutional settings, from regional idiosyncrasies assimilated to traditional ways of building to the dichotomies of planned and unplanned cities, shaped through top-down agencies or as chance-grown arrangements \cite{kostof1991city}. %(Kostof, 1991).
However, can local cultures actually leave traces in urban space? 
Despite its persistence in the urban imagination, the problem of how built environments might embody specific cultural identities seems yet to be fully addressed in urban morphology.
To begin with, there is an “evident lack of a quantitatively rigorous, comprehensive and systematic framework for the analysis of urban form” \cite{venerandi2017form}.
In this sense, historically- and culturally-informed quantitative methods are essential for uncovering forms and patterns resulting from city organization processes \cite{boeing2019urban}. %(Boeing, 2019).

In this paper, we look closely into that assumption, and address the question of %\textcolor{blue}{
whether cities find distinct regional characteristics as material forms and cultural milieu, or take on physically specific forms under certain cultural conditions \cite{hillier1989architecture}. % (Hillier, 1989).
%“we can begin to understand why it is that cities take on such different forms in different social and cultural conditions: [...] why there are administrative cities and commercial cities, which are physically as well as sociologically distinct from each other; and why there are Islamic cities and Italian cities, each with their distinct characteristics as material forms and as cultural milieux” (Hillier, 1989:6).
This implies examining the existence of contextualised ways of shaping cities -- and features that might transcend context. 
%These questions invite us to take the step \textit{from form to information} and investigate whether urban form could embody information of any kind. 
%In fact, there is a established field of work dealing with spatial (or environmental) information. 
%We shall focus on a specific kind of environmental information: the information possibly embodied in physical space.
%North American cities frequently find regular, gridiron arrangements 
%as do the historical core of cities founded by the Spanish in America; while cities in Latin America seem to find common features in spontaneously grown ‘organic’ patterns -- apparently expressing different principles of social organization and institutionalised practices.
%European cities find differences between Northern and Southern cities.
%We shall examine built form configurations in cities from different regions of the world.
We shall do so approaching the spatial configurations of the built environment as a proxy of urban culture, looking into the very constituents of urban form.
Differently from emphases on street networks \cite{hillier1993natural, porta2006network}, % (e.g. Hillier et al., 1993; Porta et al., 2006)
our approach focuses on the elementary components shaping the tangible spaces of cities: buildings and how they are aggregated in complexes of built form.
It also means taking into account a feature that seems to differentiate cities from non-urban settlements: the systems of built forms arranged in urban blocks. 
Closely related to systems of streets and open spaces, the urban block has become emblematic, uniquely defining the form of cities in urban societies emerged in regions and cultures seemingly with no contact with one other \cite{netto2017social}. % (Netto, 2017).

%We need approaches able to (a) find out whether there really are regional spatial cultures, (b) verify the classic dichotomy of the planned versus unplanned, and (c) assess whether cities from different regions fall into those categories, or what lies in between them.
%Could a concept of spatial information help us do that? 
%For that, first culturally relevant information would actually have to be encoded in the physical components of cities, i.e. buildings and how they are positioned in urban space and aggregated into blocks, giving rise to the structures we see in cities. 
% This means seeing the elementary forms of spatial aggregation through the lens of information. 
%Second, we need a measure of information able to be applied to configurations of built form.

We will look into 45 cities around the world and measure their spatial configurations to assess differences and similarities between them. In order to do so, we shall lay down an approach based on Shannon's \cite{shannon1948mathematical} measure of information and entropy.
We will argue that Shannon's measure is particularly suited for the task of capturing amounts of information related to randomness and order in configurations of built form.
Our approach takes the following steps:

% FOR NEW FIGURE [of course something similar can be said of the passage from 2D to 3D structures: a same 2D building footprint may have an infinite number of 3D shapes, from regular to irregular ones. So the ideal is to deal with 3D form. This is computationally costly, and we shall pursue that venue in a future work.]
\begin{itemize}
\item Inquire into built form as ‘spatial culture'. 
\item Propose a measure of configuration of built form based on Shannon's entropy.
\item Apply this measure to examine cities of different regions of the world.
\item Finally, use the results as a similarity measure to compare and cluster the studied cities, as ‘information signatures’ potentially associated with specific regions or spatial cultures.
%\item Verify whether the measure of information in urban form can grasp cultural differences.
\end{itemize}

%Finding consistencies in specific sets of configurations would be the first step in hypothesising causalities between information, ways of arranging built form and cultural identity. So let us first look into the problem of culture and space.

%We are going to use:
%data-driven urban morphology
%‘computational geometry’
%information theory and entropy measures applied to two-dimensional cellular arrangements.
%modeling basic spatial data to capture differences between configurations - grasped as traces of information present at the smaller scales of cellular arrangements: from blocks of two to twenty cells [check]

\section{Does culture leave traces in urban space?}
%Let us explore the possibility of cultural differences in urban form.
‘Culture’ was famously described by Raymond Williams \cite{williams1995sociology} %(1981) 
as one of the most complex words in the English language, an elusive phenomenon notoriously difficult to conceptualise and frequently challenged as an explanatory category \cite{mitchell2000cultural, gregory2011dictionary}.
% (Mitchell, 2000; Barnett, 2009). 
We use the term not to refer to an ‘independent entity’ with explanatory force \cite{duncan1980superorganic} %(Duncan, 1980) 
but as an ongoing process involving the practices, works and products of human activity situated in time and place. 
Therefore, ‘culture' is embedded in material contexts and social frameworks, and relates to institutions and institutionalised behaviours, values, meanings and orientations, and capacities for self-regulation \cite{bennett1998culture}. % (Bennett, 1998). 
Such processual notion of culture also as a field of action \cite{hastrup2004action} %(Hastrup, 2004) 
takes into account forms of self-organisation and coordination between agencies in material production.
%In this sense, there are material and contextual aspects to culture that we are willing to address.
In this sense, we wish to explore urban form as an expression of cultural systems as inherently material processes, 
aware of potential contingencies that must be considered in empirical analysis. 
%we also wish to avoid environmental determinism. We are certainly not implying that built form has particular effects over a ‘culture’, but as an expression of culture. The extent of that relation is of course very much an open question in urban studies and geography (Barnett, 2009).

One of the under-examined assumptions about the connections between society and urban form is that the latter may somehow express cultural identities that constitute the former. 
%For him, we need to know more of societal behaviours and customs to decode the spaces around us. 
Conzen \cite{conzen1960alnwick} %(1960) 
was a pioneer in studying patterns of change in urban form in relation to changes in the economic, social, and technological milieu, proposing a cyclical nature of the development of urban form \cite{rashid2017geometry}. %(Rashid, 2017). 
Going a step further, Aldo Rossi \cite{rossi1982architecture} %(1966) 
argued that the material form of the city is intrinsic to its sociological and cultural reality. Later on, Hillier \cite{hillier1989architecture} addressed the possibility of cities of different cultural types embodying different spatial identities. His analytical approach allowed him to claim that human societies order their built environment to construct a `spatial culture', a `distinctive way of ordering space'. Cities take on different forms in different cultural conditions in non-contingent ways, as spatial arrangements shape the field of encounters that animate different social cultures. Physical space is systematically ordered to reproduce culturally-specific patterns of social behaviour based on co-presence as a principle for ordering social relations.

Recent discussions have enriched this construct by approaching a spatial culture as “a fundamentally performative and temporal process” \cite[p.xxiv]{griffiths2016spatial}
%e' a pagina? Pode por na citação na bibliografia...}.
%(Griffiths and Von Lunen, 2016: xxiv).
Focusing on “questions of cultural specificity in the formation of space”, these works assess how culture affects spatial formation, and the possibility of “encoding and transmitting social and cultural information” in urban space \cite{koch2016spatial}. % (Koch, 2016: i).
Contingencies are added by the possibility that “different cultures invest differently in space, be it in regards to what is manifested, or to what extent society is manifested through built form” \cite{koch2016spatial} (p.i). %(Koch, 2016: i). 
Difficulties are also of an epistemological nature, since the space-culture relation may be indivisible analytically \cite{palaiologou2016reclaiming}, %(Palaiologou et al, 2016: 28), 
and therefore hard to be scrutinised. 
%Since Peponis's (1985) reference to the architectural scale, the term ‘spatial cultures' has received growing attention in urban configurational studies. 
%The first exposition of the term in relation to cities is found in Hillier (1989; see Griffiths and Von Lunen, 2016).

In a careful opposition to some of these views, the late urban historian Spiro Kostof \cite{kostof1991city} was suspicious of the belief that buildings and city-forms fully embody recognisable idiosyncrasies enough to be medium of cultural expression. Even though his works relate processes like `reading', `encoding' and `information' to culture, he claimed that a same urban form does not express an invariable human content. Despite this position, the quest for underlying explanations for systematic differences in urban form has led to the idea that physical patterns encapsulate an extra-physical reality, %[25], 
as different social 
and cultural agencies 
are seen to shape physical space. 
These agencies can range from 
%actions of people producing space according to 
tradition and
custom, material requirements of interaction, associations with socially shared symbols or principles of societal organisation. 
For instance, cities with irregular physical patterns are thought to be the result of development left entirely to individuals, as bottom-up processes leading to the random ways of the unplanned city. In turn, top-down processes triggered by governing bodies would be able to guide the organisation of urban land and built form, leading to %\textcolor{green}{
uniformly ordered
%patterned
cities [Castagnoli in \cite{kostof1991city}, p.43)]. %Kostof, 1991:

%However,  if there is a presence of cultural idiosyncrasies encoded in space enough to distinguish specific identities, 
%How does the idea of urban form expressing culture connect with regional consistencies, and different planning cultures?
Some studies looked into spatial features, logics or organising principles in comparative studies of cities consistent with distinct regions.
%to some extent associated with versions of `planned' and `unplanned' spatial cultures.
To be sure, most of these works deal with street networks rather than built form systems.
Medeiro's \cite{medeiros2013urbis} topological analysis of betweenness centrality and depth in the street networks of 164 cities in different parts of the world identified regional differences. %with typical accessibility levels. 
For instance, American and Canadian cities appear prominently with the highest levels of accessibility, as opposed to Brazilian cities in South America, the most spatially segregated.
%\textcolor{green}{
Louf and Barthelemy \cite{louf2014typology} 
searched for the ‘fingerprints’ of cities 
%using the hierarchical clustering method to 
analysing the distribution of blocks extracted from street networks of 131 city centres.
Their classification scheme is based on information about the area and a simplified proxy for the shape of blocks.
%Their typology is essentially based on the ratio of urban block area and the area of the circumscribed circle.
%Since different blocks can offer a same ratio, this measure is rather an approximation of form: it does not preserve the actual shape of blocks or deal with built form. 
%Furthermore, the method does not grasp information encoded in relations between blocks, as a cellular approach to building footprints can.
%This interesting approach describes shapes based on a form  factor }
The method identifies that nearly two thirds of American cities in their sample are structurally different from European cities.
%finding large ‘families of cities’, e.g. most European cities and American cities fall in their own sub-category.
Rashid \cite{rashid2017geometry} carried on a comparative study of urban morphology in 104 cities in six continents. 
%(34 cities from North America, 23 from Europe, 7 from Australia and Oceania, 9 from South and Central America, 16 from Asia, and 15 cities from Africa). 
Using uni-variate statistics of data for 44 spatial measures of street configurations and basic geometric measures of built form, like block perimeters and areas, he found limited differences between downtown areas in developed and developing countries.
Furthermore, disaggregated measures of geometry of urban layouts have little power to describe actual form (say, of blocks) and do not grasp information encoded in relations between components of built form. %i.e. buildings or cellular arrangements.

In turn, our configurational approach focuses on buildings as the elementary components shaping cities, and how they are aggregated in combinations and complexes. 
By looking into frequencies of  
cellular arrangements representing buildings in selected cities, we wish to understand \textit{if} and \textit{to what extent} their configurations can be seen as particular cultural features, regardless of whether these features are %`codes' 
intentionally embodied in urban space.
Recognising that urban structures are different around the word, and approaching spatial configuration as a proxy of urban culture, we attempt to measure such configurations to assess their differences and similarities. 
For that, we shall explore Shannon's view of `information' and `entropy' to investigate whether spatial cultures entail `distinctive ways of ordering space', as Hillier suggests.

\section{Information and entropy in physical spatial systems}

%Of course the idea that cities embody information is well known in urban theory. It is at the heart of Kevin Lynch’s \cite{lynch1960image} pioneering work in 1960 on spatial elements guiding navigation in the environment, along with memory and representation, even though he did not quite use the term ‘information’. In the 1980s, Amos Rapoport \cite{rapoport1982meaning} (page 19) explicitly asserted, “physical elements of the environment do encode information that people decode”. 
%Hillier and Hanson \cite{hillier1984social} thought of non-representational meaning embedded in physical configurations, as patterns guiding way-finding correlated with patterns of co-presence. 
%More recently, 

A number of works have explored information and entropy measures in relation to urban systems, beginning with Wilson's \cite{wilson2011entropy} pioneering study of utility-maximising systems in 1970. The entropy-maximising paradigm was frequently used to derive model formulations for spatial interactions and urban distributions, microeconomic behaviour and input-output analysis \cite{batten1981entropy}. %(Batten, 1981). 
Batty developed a number of studies on entropy in spatial aggregations and interaction since the early 1970s \cite{batty1972entropy, batty1974spatial, batty1976entropy}.
More recently, Batty et al. \cite{batty2014entropy} 
proposed a measure of complexity based on Shannon information able to grasp the complexity of cities as they vary in scale, size and spatial distribution of population, dealing with spatial entropy related to the distribution of information, and with information density related to city size.
Other approaches used modifications of Shannon entropy and information-theoretical metrics as methods to capture, quantify and group similar two-dimensional spatial patterns in landscape ecology, including efforts towards a universal classification of configuration types in a linear sequence according to increasing values \cite{claramunt2012towards, altieri2018new, nowosad2019information}.
%(Nowosad and Stepinski, 2019).
%(Claramunt, 2012; Altieri et al., 2018) 

\begin{figure}[h]
    \centering
    \includegraphics[width=1.0\textwidth]{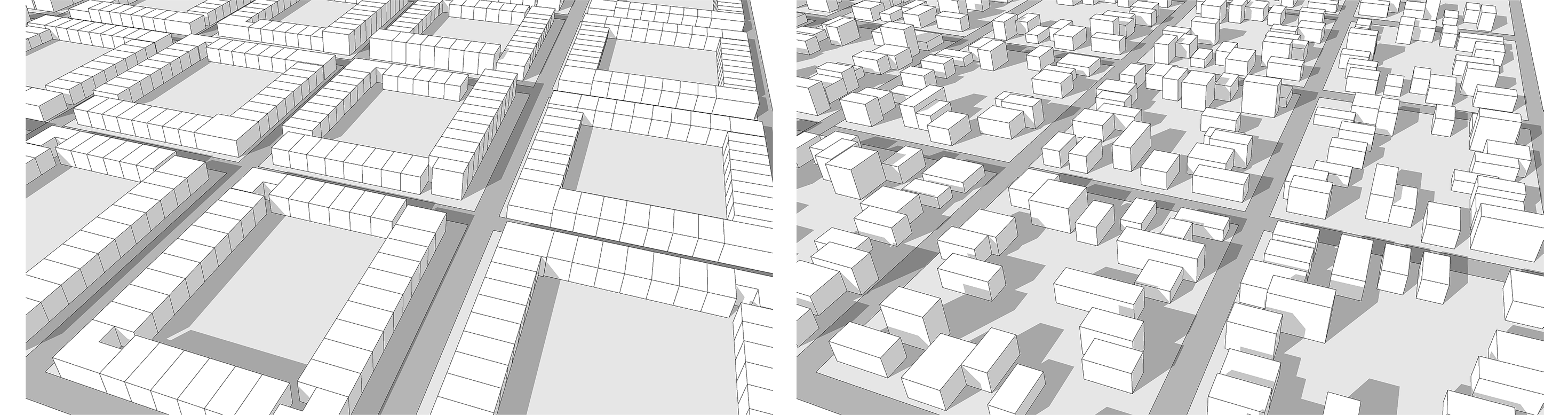}
    \caption{\small A same street network can support very different built form systems.
    }
    \label{fig:Fig 1_built form variation_h.png}
\end{figure}

Entropy measures have also been applied to purely urban morphological problems, namely in street network analysis. Gudmundsson and Mohajeri \cite{gudmundsson2013entropy} developed a method based on Shannon's entropy to measure angular variation between streets, applied to 41 British cities.
Boeing \cite{boeing2019urban} applied Gudmundsson and Mohajeri's method to analyse 100 cities around the world focusing on street networks downloaded from Open Street Maps (OSM). 
However interesting as morphological approaches, these applications do not seek to uncover spatial information patterns, focusing instead on entropy as a measure of variation in street angles and lengths. %\textcolor{green}{
The entropy measure of the distribution of crossing angles does not necessary capture the global degree of order/disorder of street networks. 
Even if they do, %} it happens, The entropy measure applied to 
street orientation does not describe spatial configurations in a relational sense.
%Also, resulting values are concentrated in extremes, suggesting that the measure is not so sensitive.
%\textcolor{green}{
These considerations are reflected by the fact that the resulting values of the measure applied are concentrated in the extremes, suggesting that it is not so sensitive. %} 
Furthermore, entropy measures applied to street orientation are not a comprehensive morphological approach since they not take into account entropy in built form. 
In other words, they ignore discrepancies between levels of order in street networks and built form systems. 
Cities can be physically disordered even if their street networks are perfectly ordered.
As Kostof \cite{kostof1991city} %(1991:44) 
put it, “[s]treets that read as straight and uniform on the city plan may be compromised  by the capricious behavior of the bordering masses” (p.44).
In short, we can have low entropy in street orientation, yet highly disordered morphological structures (figure \ref{fig:Fig 1_built form variation_h.png}). 
%(some could say the same about the 2D-3D difference: a perfectly ordered 2D city could be disordered in the third dimension, e.g. Manhattan)

More comprehensively, Haken and Portugali \cite{haken2003face, haken2014information} focused on how the built environment actually embodies information. They explore Shannon information quantitatively in connection with Haken's synergetic qualitative approach to semantic information, in order to empirically assess how basic cellular arrangements and categorisations of building facades convey different amounts of information.
Finally, other approaches to spatial information have adopted different measures of entropy, distribution of spatial co-occurrences, or information density to assess the amount of redundancy and grouping related to cognitive efforts to extract task relevant information from the built environment \cite{woodruff1998constant,rosenholtz2007measuring}. 
In turn, our approach will explore Shannon  entropy %information 
to measure levels of randomness and disorder in physical space, namely in cellular arrangements of built form. %capturing amounts of information related to configurations of built form.
We shall look into the possibility that consistent differences between cities can be perceived at this scale, 
and that cultures and regions find different ways of ordering such configurations, which may be captured by this measure.

%Exploring information in physical space, our approach focuses on the role of order in the arrangement of buildings.
%In that sense, we analytically suspend semantic contents of space produced by social activity to focus purely on physical configuration.

\section{Analysing built form systems}

Our first procedure involves a reduction of urban form to two-dimensional arrangements based on building footprints.
Since Giambattista Nolli’s 1748 Map of Rome, the figure/ground diagrams have become a classic methodological resource in urban studies, showing built/unbuilt distinctions. For instance, Nineteenth century scholar Camillo Sitte \cite{sitte1979art} represented public buildings and the ordinary fabric of the city exploring such diagrams. 
%Giesler [10] has used such maps for representing building fabric and land utilization in urban geography.
More recently% in urban design
, Rowe and Koetter \cite{rowe1983collage} have described the theoretical significance of the figure-ground map or `Nolli map'
%Habraken [12] has used it to create built and unbuilt and thematic (typical) and non-thematic space (atypical) 
\cite{rashid2017geometry, hwang2005heuristic, verstegen2013giambattista}.
%(Rashid, 2017; see Hwang and Koile, 2005; Verstegen and Ceen, 2013). 
The figure-ground diagram provides a spatial data-driven method to analyse and study the urban form and circulation networks that structure human activities and social relations \cite{boeing2019urban}. %(Boeing, 2019).

\begin{figure}[h]
    \centering
    \includegraphics[width=0.8\textwidth]{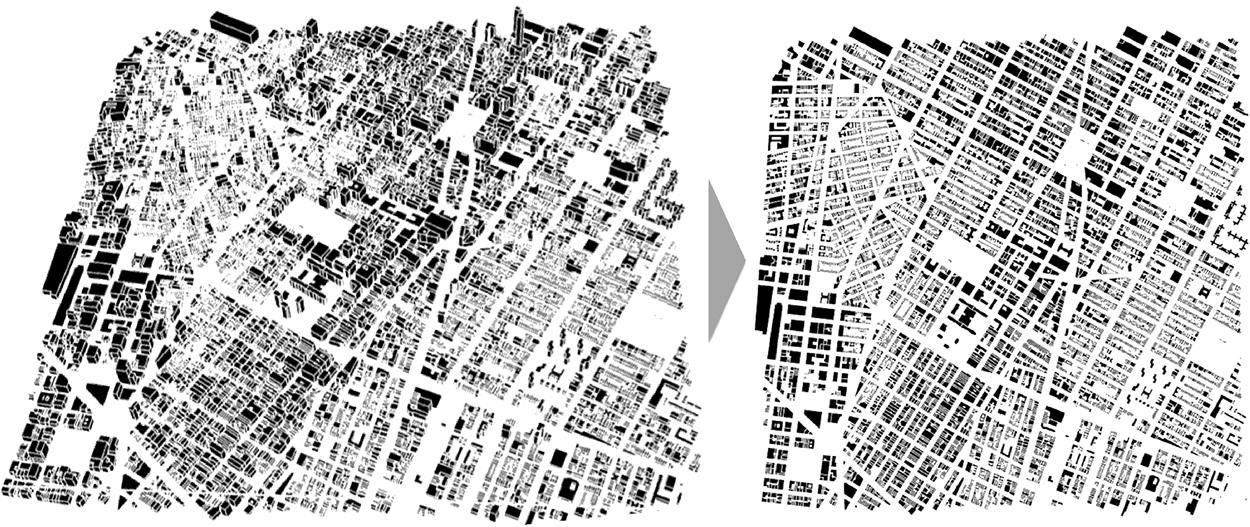}
    \caption{\small Reduction from 3D to 2D configurations of built form (Manhattan, NYC).
    }
    \label{fig:3D_2D_config2.png}
\end{figure}

Our second procedure looks into different cellular arrangements of built form and attempts to characterize their configurations.
%\textcolor{blue}{
We do so analysing the probability distribution of built form configurations, by estimating the Shannon entropy \cite{shannon1948mathematical} of Nolli maps of different cities of the world.
%}
%\textcolor{green}{ We do so estimating the Shannon entropy \cite{shannon1948mathematical} of the Nolli maps of different cities of the world.}
%to measure the probability of different combinations of cells, assessing their frequency in different cities of the world. 
Of course, this has to do with the level of randomness in the cellular arrangements of built form in cities.
By analysing cellular arrangements, we capture the structures of urban blocks in relation to the open spaces of streets and public squares. 
%\textcolor{red}{Assuming that the measure of entropy is able to capture the probability of configurations, this brings some methodological implications:
%\begin{itemize}
%\item Is the measure of entropy enough to capture information embodied in urban form?
%\item Is the scale of built form cells enough to encode information specific enough to be differ from that found in other cities? That is, can we capture specific spatial cultures by looking at the local arrangement of built cells? Can a physical local measure of information capture qualitative differences - if they exist?
%\end{itemize}}
Indeed, the layout of the environment encodes more information than two-dimensional configurations can express.
However, we opted for an analytic approach able to sufficiently describe differences in built form -- hence the reduction of 3-dimensions (3D) urban form to 2-dimensions (2D) cellular aggregations (figure \ref{fig:3D_2D_config2.png}).

%further develop and apply a new approach \cite{netto2018cities} to assess urban configurations as environmental information. 

We characterise the spatial information encoded in two-dimensional configurations of buildings in the following terms. 
As mentioned, information will be quantified measuring Shannon entropy, operationally estimated by looking at the sequence of bits 1 and 0 representing built form cells and open space cells within sections of cities.
Theoretically, this corresponds to measuring the Shannon 
entropy of a 2D symbolic sequence of 1 and 0. 
In this context, information finds a precise meaning: the entropy of the sequence, a measure of the surprise a source that produces the sequence causes in the observer \cite{shannon1948mathematical}.
Physical arrangements characterised by higher levels of randomness, uncertainty or unpredictability are associated with high entropy. 
In contrast, the presence of regularities and patterns in urban structures corresponds to lower entropy, which means a higher predictability.

The next step involves the preparation of our set of empirical cases, and the conversion of city maps into Nolli maps. 
We selected cities for their importance in their region or country. 
% based on the availability of data and \textcolor{red}{with a size compatible} with our methodological requirements.
%The first item has brought us to some well-known cities as emblematic cases. 
Selection also had to take into account the availability of information on built form. 
Many cities, particularly in Latin America, Africa and Asia, have incomplete information regarding building footprints, i.e. their precise location, position and form. %For instance, major cities in South America like Lima and Bogota could not be included, \textcolor{blue}{and we could only include Lagos in Nigeria among African cities.}

For methodological reasons, 
we selected areas within these cities for the application of our measure.
This selection procedure follows two critical considerations. 
The first and most important one observes that it is interesting to decouple the analysis of urban structures between small-scale, detailed and denser urban areas, and large-scale regional and peripheral urban areas.  
In fact, the two areas are different, and for this reason, they can be naturally described using different methodologies. 
The first small-scale urban area is defined by specific features such as buildings and urban blocks,
which introduce typical characteristic scales.
% \textcolor{green}{
This means that there are some well-defined scales related to the distance above which configurations loose their correlations. These characteristic scales 
define sub-systems characterised by typical local patterns (urban blocks, individual buildings and possible neighborhoods).
Here, human %intervention shapes 
action is the principal vector
defining shapes and patterns  
which generally appear in a stratified form, like the ones we see in older and traditional central areas. 
In turn, large-scale regional and peripheral urban areas are likely to include sparse occupation, frequently with a scale-free character. 
This means that the characteristics of their patterns  are independent of the scale we fix for analysing them. 
Looking at different scales, the underlying structure remains the same.
%This means that they lack a characteristic scale
%and may include 
In these regions, physical features
linked to topography, geographical formations and barriers (e.g.. water bodies, mountains, and valleys), along with the presence of very large infrastructures (e.g. highways) might play %the most 
relevant roles in the definition of the spatial patterns.
In this work, we will focus only on small-scale areas with dense %and continuous 
urban form.

The second consideration takes into account that our method is well fitted for estimating entropy for dense and 
continuous %and homogeneous 
urban areas. % \textcolor{green}{
Fixing the density of built form cells %per total cells including open spaces 
allows us to obtain results independent from this parameter. 
The high continuity and homogeneity
of built form allows us to use a specific extrapolation technique that will %implement 
prove useful for estimating the entropy of our 2D symbolic sequences.
For these reasons, the selection of  sections was based on the identification of dense areas, with a high spatial continuity in the fabric of built form. We will consider %able to satisfy
occupation rates close to 50\%, which means avoiding large empty areas or rarefied patterns of urbanisation.

\begin{figure}[h]
    \centering
    \includegraphics[width=1\textwidth]{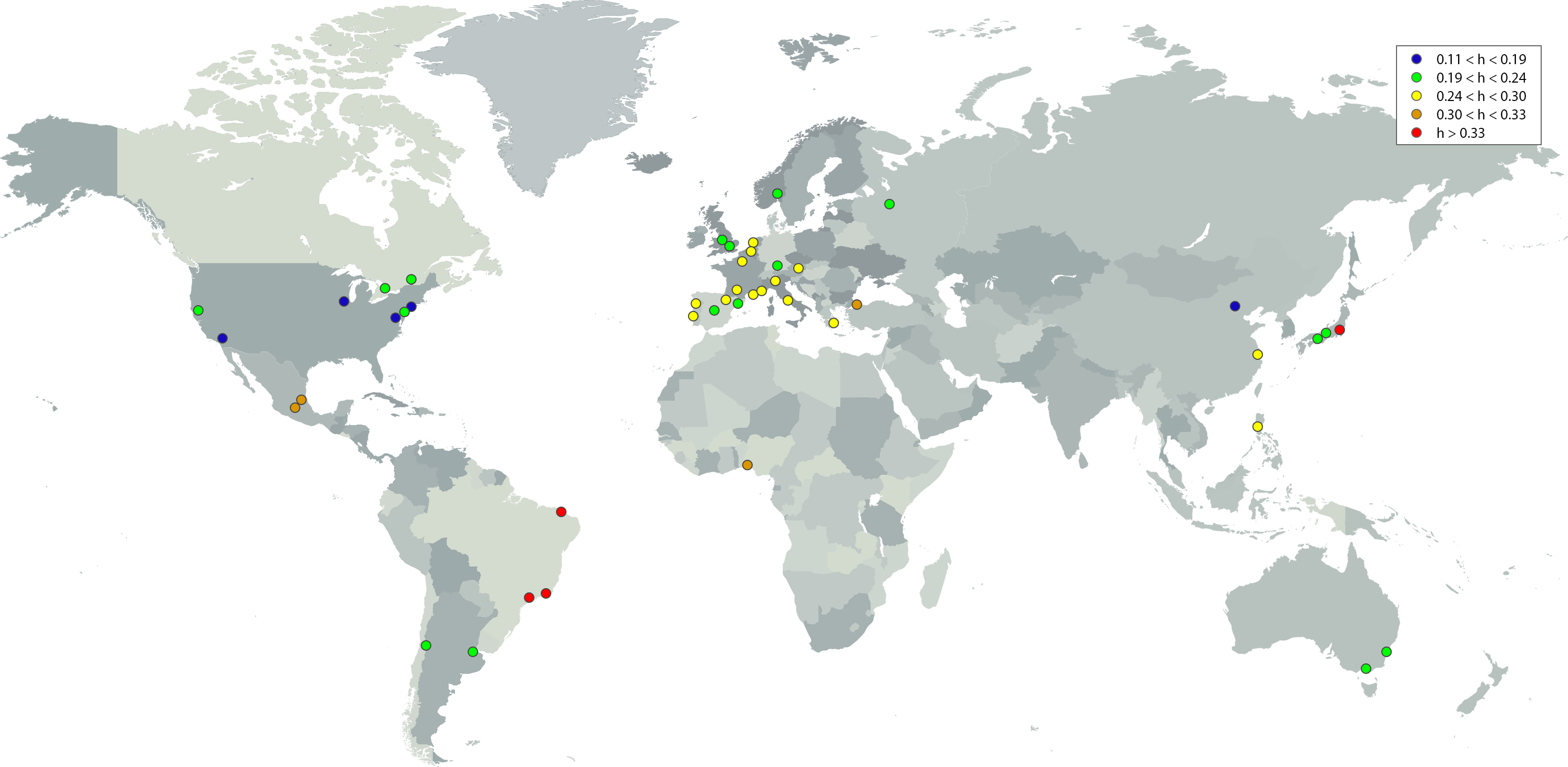}
    \caption{\small Location of 45 cities in our sample. Colours show amounts of Shannon entropy found by our method in cellular configurations of built form, from blue (low entropy) to red (high entropy).
    }
    \label{fig:Fig3_new2.png}
\end{figure}

\begin{figure}[h!]
    \centering
    \includegraphics[width=0.70\textwidth]{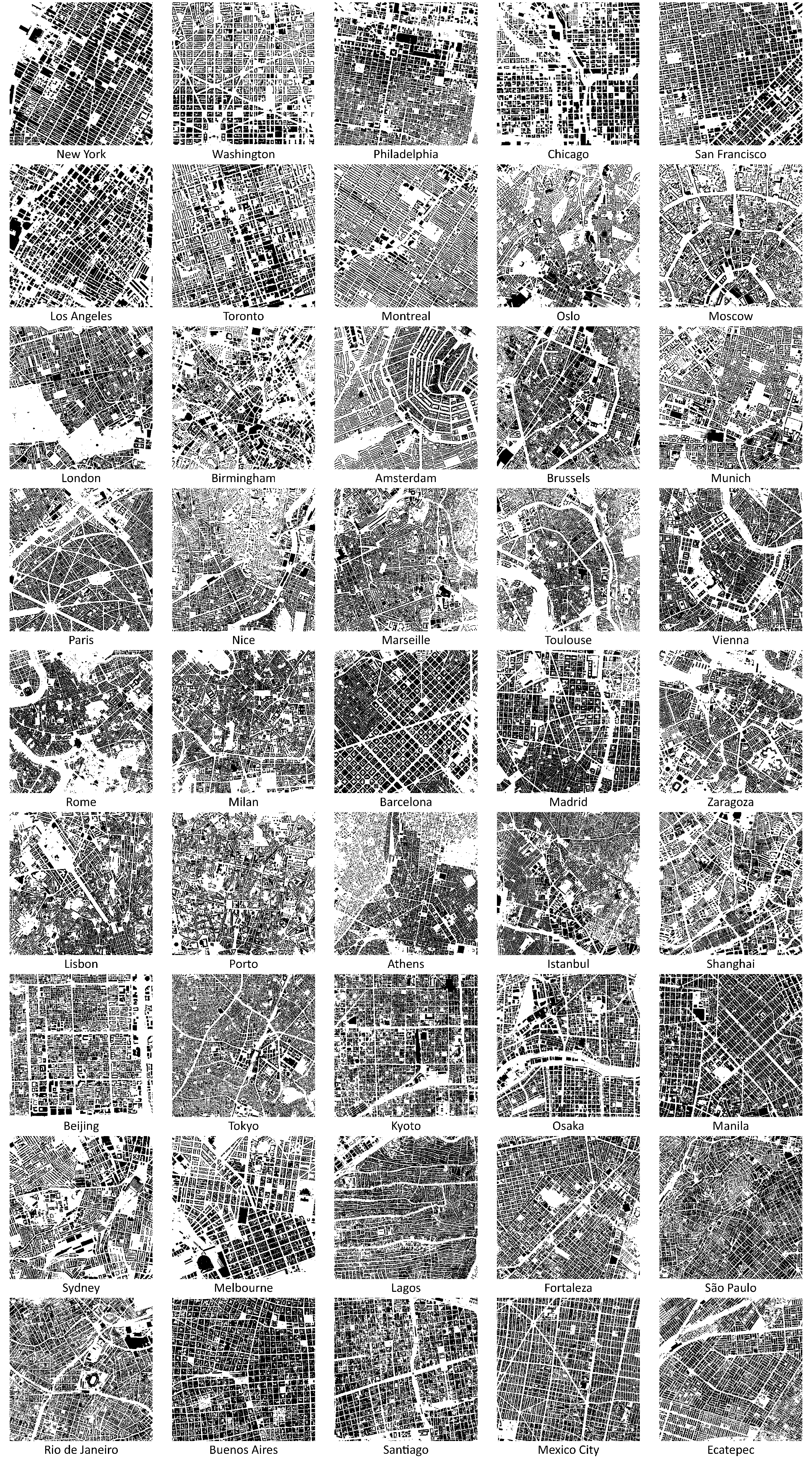}
    \caption{\small Nolli maps with building footprint distributions in downtown areas of 45 analysed cities (9,000,000 m$^2$ windows, 1,000,000 cells), extracted from Google Maps. These sections are used to compute Shannon entropy. Rotation in grids and built form systems does not affect results.
    }
    \label{fig:Figure_4_45 cities.png}
\end{figure}

We prepared our sample extracting building footprints in sections of cities from the public map repository Google Maps API. 
We tested trade-offs between resolution and availability of data for distinct scales. 
We chose geographic areas of $9,000,000$ m$^2$, which were considered sufficient for representing the general spatial characteristics of dense urban areas regarding the configuration of buildings, urban blocks and open spaces of 45 cities around the world (figure \ref{fig:Fig3_new2.png}).
Built form maps of the selected cities were then prepared and exported in high resolution, filtering layers and converting entities representing buildings into solid raster cells. 
Images underwent a re-sizing process for $1000^2$ cells and were converted to a monochrome system and then into a matrix of size $1000\times 1000$ cells with binary numerical values (figure \ref{fig:Figure_4_45 cities.png}).\\

%FIGURE: 45 CITIES_matrix

Estimation of the Shannon entropy of the considered 2D cellular arrangements uses a method commonly applied for estimating the entropy of sequences of symbols encoded in one-dimensional strings \cite{schurmann1996entropy}.
%It has been widely used for different type of data sets, from natural languages, speech analysis and behavioural sequences to deoxyribonucleic acid (DNA) and spike emissions in neurons.
%However, estimating entropy is far from trivial. 
For 1D data sets, % corresponding to one-dimensional strings,
the method consists of defining the block entropy of order $n$ through

\begin{equation}
H_n=-\sum_k  p_n(k) \log_{2}[p_n(k)],
\label{1entropy}
\end{equation}
where blocks are string segments of size $n$, and the  sum runs over all the $k$ possible $n$-blocks.
Equation~(\ref{1entropy}) corresponds to the Shannon entropy of the probability distribution $p_n(k)$.
The Shannon entropy %density
of the considered system (the whole 1D string) \cite{schurmann1996entropy, lesne2009entropy}, which we indicate with $h$, is obtained from the following limit:

\begin{equation}
h=\lim_{n \to \infty} H_n/n,
\label{hentropy}
\end{equation}

which measures the average amount of randomness per symbol that persists after all correlations and constraints are taken into account.
%\textcolor{green}{
The above limit exists for all spatial-translation invariant systems, as demonstrated in \cite{cover1991elements}.
%T. M. Cover and J. A. Thomas. Elements of Information Theory. John Wiley & Sons, Inc., 1991.
More details about this method can be found in \cite{schurmann1996entropy,lesne2009entropy}.

\begin{figure}[ht]
    \centering
    \includegraphics[width=0.8\textwidth]{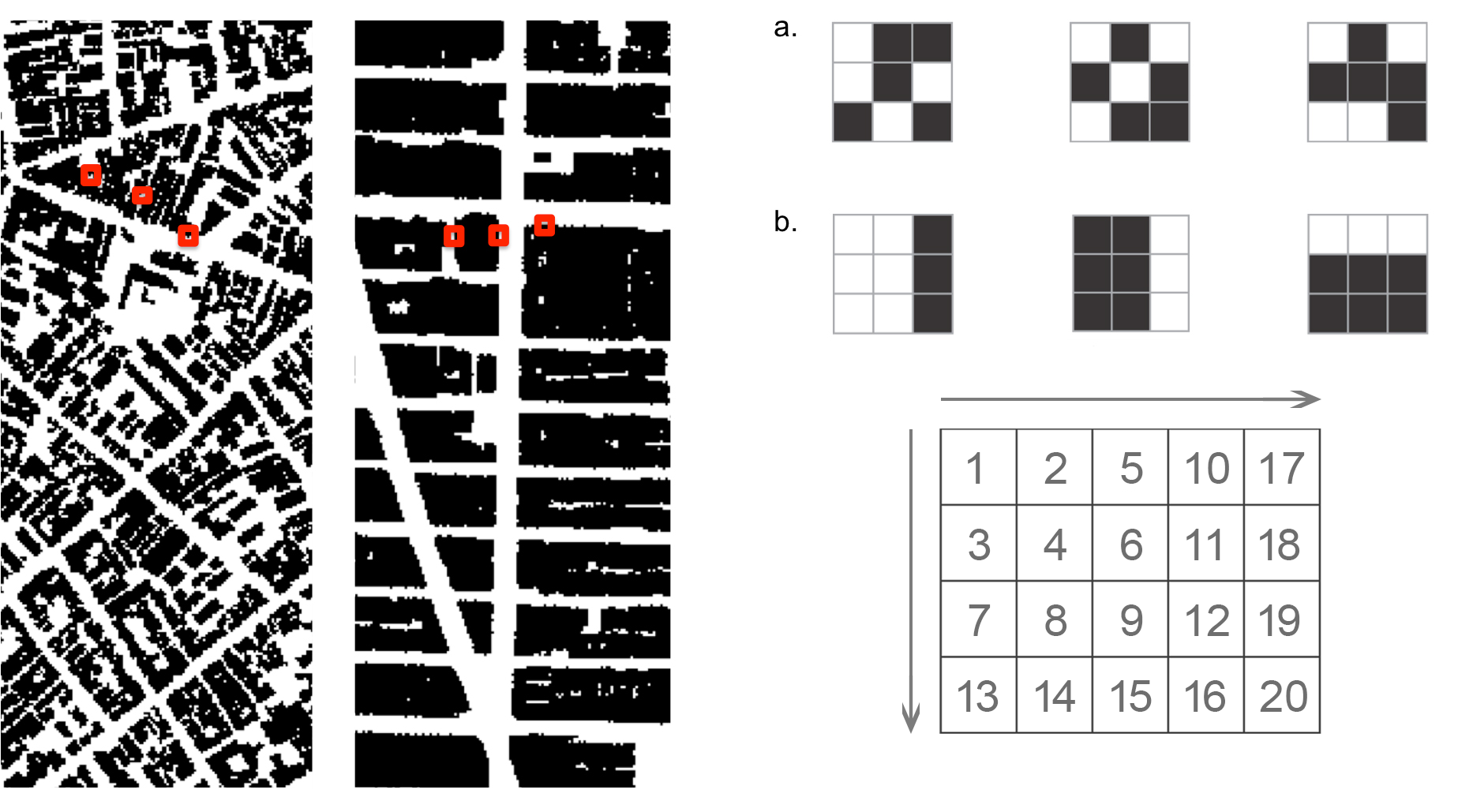}
    \caption{
    Areas in Rio de Janeiro and Manhattan, NYC (left). 
    Examples of blocks with nine cells shown in red are amplified on the right. 
    Configurations of the type (a) show great variation, like those found in Rio, while the type (b) shows regular arrangements frequently found in Manhattan.
    Blocks are constructed following the fixed path represented on the bottom right. Numbers indicate the order in which cells are added to blocks. The first block of size 1 corresponds to cell 1. Neighbouring cells are added in the corresponding order.
    %Essentially, p_n(k) in Equation (1) accounts for the number of times every possible configuration k appears in the map for a block of size n.
    %\textcolor{blue}{Block sizes are analysed up to $n\approx 20$. 
    Nolli maps are scanned with this set of different cell blocks.}
    \label{fig:blocks+20cells2.png}
\end{figure}

This approach can be generalized to sequences of symbols in two dimensions, which correspond to our situation. 
%For this method can be applied to our problem once 
We have to define the $n$-blocks for a two-dimensional matrix \cite{feldman2003structural}. %In this two-dimensional context, 
The most intuitive idea is to consider a block of size $n$ as a square which contains $n$ cells. 
To obtain the sequence of $H_{n}$ also for $n$ values that do not correspond to squares, we considered blocks that interpolate perfect squares, as described in figure \ref{fig:blocks+20cells2.png}.
Note that there is no unique natural way to scan a 2D matrix. 
We tested our approach for different reasonable forms of constructing the blocks, and the use of different paths does not seem to significantly influence the estimation of $H_n$ for the considered data set.

Equation \ref{hentropy} gives precisely the entropy for a theoretical infinite set of data.
In real situations, where the data set is finite, our method estimates the probabilities of distinct arrangements of cells within blocks up to a certain size $n$, counting their frequencies. %, and then estimates the limit.
For example, for $H_1$, it is sufficient to have knowledge of the symbol distribution $p_1(2)$, which is approximated by the frequency of 0 and 1 present in the data set. 
It is important to note that it is common to find in the literature of image processing, urban studies and ecological landscapes approaches that perform some entropy based analysis measuring our $H_1$ or, at best, our $H_2$. 
Unfortunately, sometimes these quantities are wrongly referred to as the Shannon entropy of the system, %\textcolor{green}{
which, in contrast, is our $h$.

If our data were a purely random set, $h$ would coincide with $H_1$, and $p_1(2)$ would give a full account of the spatial configuration. 
This is obviously not true for urban situations, where evident structures and strong long-range correlations are present. In this case, estimating entropy is a difficult task, as taking correlations into account means computing $H_n$ for a large $n$. 
In fact, the estimation of $h$ is good when the spatial range of correlations %\textcolor{red}{and memory} 
is smaller than the maximum size of the block entropy we are able to compute. This estimation can be rendered difficult because of the exponential increase in the number of distinct cells arrangements in blocks with $n$. %($k=2^n$). 
%For instance, there are 512 different configurations for blocks with only nine cells. 
%Difficulties in capturing longer correlations lead to the overestimation of h. This is the case if sufficient care is used in estimating each $H_n$. Otherwise, as strong fluctuations are already present for moderate block lengths $n$, the estimates $H_n$ are usually underestimated. These two concurrent effects may jeopardize the estimation of entropy.
%Theoretically, the value of $h$ should not depend on orientation or rotational transformation of the cellular grid. 
When working with two symbols, as in our case, the estimation of $H_n$ becomes not reasonable when $2^{n}\approx N$, where $N$ is the number of elements in our data set. 
Thus, in our case, this condition is verified for $n\approx 20$.
Even if this is a rough evaluation, 
it reasonably fixes the maximum size of the blocks that can be %included in the analysis, 
investigated with sufficient statistical quality \cite{lesne2009entropy}.
%it is a reasonable divide between good and bad statistics 
%, which means blocks of squares with a linear length smaller than 5 cells, which corresponds to $15$  meters ($m$).
%It follows that, if a city has relevant internal correlations larger than $15$ $m$, the entropy estimation using block entropy will be over-estimated.
The limit taken in equation \ref{hentropy}
can be empirically obtained fitting the 
set of $H_n/n$ points with an appropriate function
and then taking its limit for $n \to \infty$.

We found heuristically that, for all examined cases, the following ansatz provides an excellent fit:
\begin{equation}
H_n/n \approx a+b/n^c, \qquad  b,c>0.
\label{fitting}
\end{equation}
%Even if we observed that the convergence is relatively slow, 
The fitted value of $a$ gives a reasonable extrapolation of the Shannon Entropy $h$.\\

Considering our database of 45 cities from North America, Europe, Asia, Oceania, Africa and South America, our goal is to develop a classification scheme based on the similarities and differences between the entropy levels
of the sampled cities. In this sense, the next step consists in performing a proximity network analysis based on the measured entropy values, with the aim of identifying the presence of communities or clusters of cities sharing similar entropy levels. 
In short, entropy estimation
will allow us to order our pool of cities and define a classification scheme. 
This scheme may help us %in verifying the possibility of finding
find similarities %in the configuration of cities 
possibly consistent with same spatial cultures or world regions. %along with differences between cities from other cultures.

Once we  obtained the entropy $h$ for all considered cities, we can quantify the levels of similarity defining a distance  between cities $i$ and $j$ based on the values of $h$:
$d_{ij}=|h_i-h_j|$. We created a matrix of distances for the analysed cities and then defined a network where cities are nodes, and edges (links between nodes $i$ and $j$) are present only if the value of $d_{ij}$ is smaller than a fixed threshold value.
%of $0.03$, which roughly corresponds to the 99\% C.I. of the extrapolated values of $h$.
The detection of clusters displayed by this network is a
straightforward task considering the relatively small size of our data set.

We further developed the cluster analysis applying a method for constructing a dendrogram representation of the distance matrix. 
%phylogenetic-like trees to the above mentioned distance matrix. 
We used the unweighted pair group method with arithmetic mean (UPGMA). This method constructs a dendrogram that reflects the structure present in the similarity matrix, building a hierarchy of clusters. 
The algorithm used in the analysis is part of the module Bio.Phylo in the Biopython package \cite{biopython}.
%  cite: https://biopython.org/wiki/Phylo
When this approach incorporates a reliable dating of %the considered
entities, it can be used to identify %for defining
cultural phylogenies, like in the work of Barbrook et al. in the phylogenetic analysis of written texts \cite{barbrook1998phylogeny, howe2001manuscript}.
%Barbrook: Barbrook AC, Howe CJ, Blake N, Robinson P. 1998 The phylogeny of The Canterbury Tales. Nature 394,
%839. (doi:10.1038/29667); 
%Howe JC, Barbrook AC, Spencer M, Robinson P,Bordalejo B, Mooney LR. 2001 Manuscript evolution.Trends Genet. 17, 147 – 152. (doi:10.1016/S0168- 9525(00)02210-1)
This is an interesting exception.
In general, like in our case, 
%\textcolor{blue}{cultural objects are mutually involved and related between themselves}, 
cultural objects are related in an involved form between themselves, 
and
dating is a major challenge for long standing living entities, whether they are cities or languages \cite{cavalli1997genes, cavalli2001genes,benedetto2002language}.

\section{Results: proximity networks and hierarchical clustering} 
\begin{figure}[h]
\begin{center}
\includegraphics[width=0.4\textwidth]{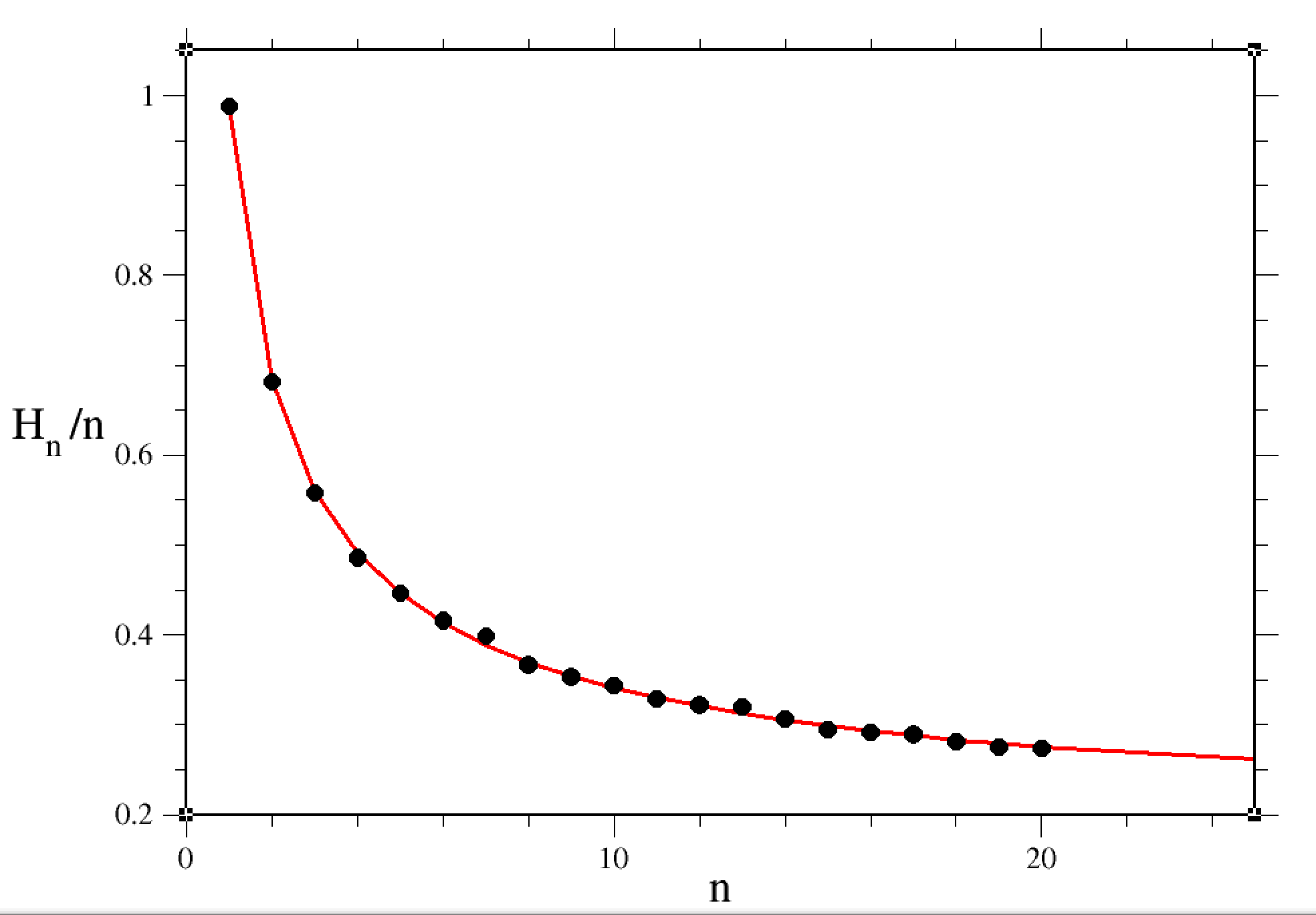}
\includegraphics[width=0.59\textwidth]{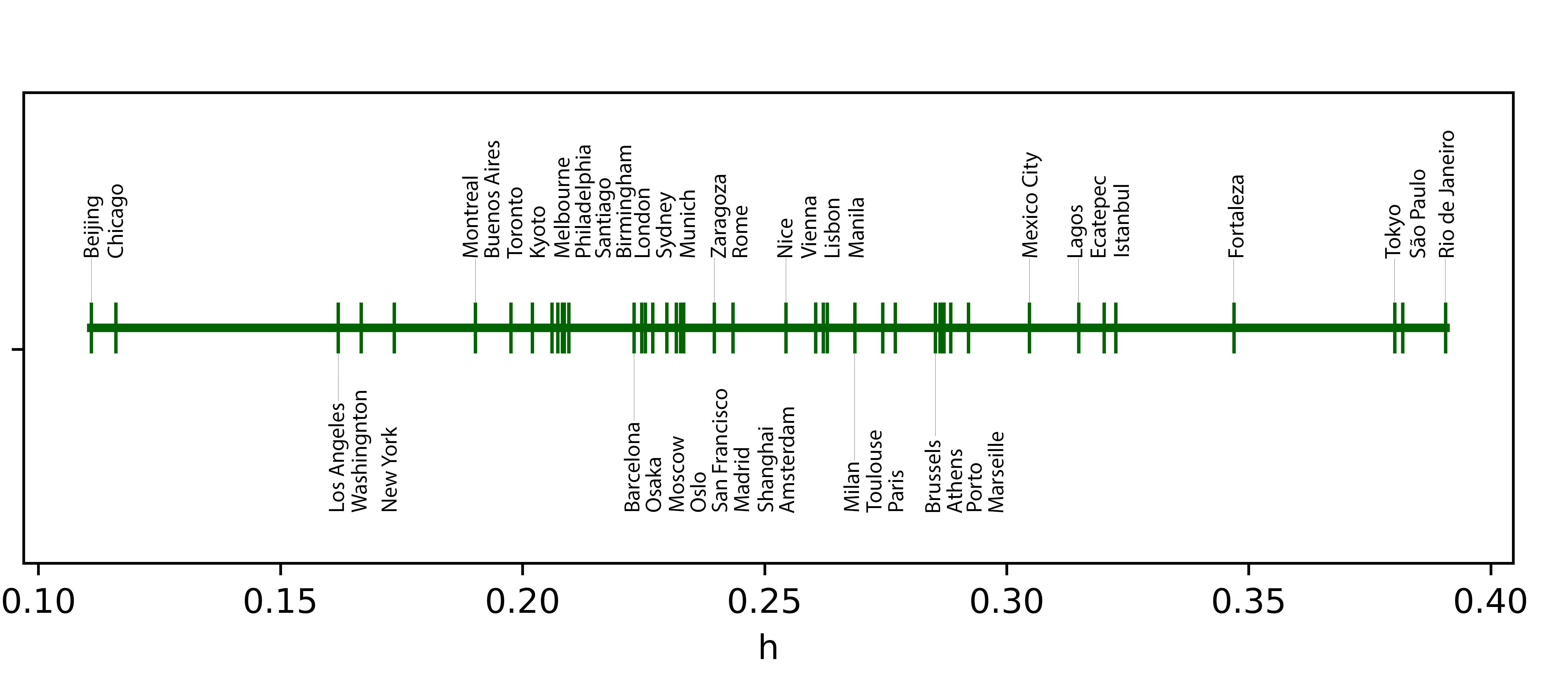}
\end{center}
\caption{\small \textbf{Left}: An example of the estimated values of $H_n/n$ for the city of Los Angeles. The continuous line represents the best fitting of our data using the function of equation \ref{fitting}. %The fitted values of $a$ give an extrapolation of the Shannon Entropy $h$ of the data set. 
All the analysed cities present a very similar behaviour.
\textbf{Right:} Estimated values of $h$ for the 45 cities under analysis.}
\label{fig:ranking_2.png}
\end{figure}

The use of the empirical functions of equation \ref{fitting} provides 
an excellent fit for all the considered cities.
The values of the parameters $c$ are contained in the interval $[0.37,0.68]$. 
These values are consistent with the entropy convergence found in written texts, where $c$ ranges from 0.4 to 0.6 \cite{ebeling1991entropy, ebeling1997prediction, ebeling1994entropy}, and with a result for a  Beethoven sonata where an exponent 0.75 was found \cite{anishchenko1994power}. 
%anishchenko: V. S. Anishchenko, W. Ebeling, and A. B. Neiman, ‘‘Power law distributions of spectral density and higher order entropies,’’ Chaos, Solitons, Fractals 4, 69–81 1994.
%ebeling: W. Ebeling and T. Poschel, ‘‘Entropy and long-range correlations in literary English,’’ Europhys. Lett. 26, 241–246 1994; W. Ebeling, ‘‘Prediction and entropy of nonlinear dynamical systems and symbolic sequences with LRO,’’ Physica D 109, 42–52 1997; W. Ebeling and G. Nicolis, ‘‘Entropy of symbolic sequences: The role of correlations,’’ Europhys. Lett. 14, 191–196 1991.
These results seem typical of language-like systems, where the presence of long-range order is characterised by a slowly decaying contribution to the asymptotics of the entropy for large $n$.
Despite the relative slow convergence,
the fine quality of the fits allows a good extrapolation of the Shannon Entropy $h$.
As an example, the results for the estimation of $H_n$/$n$ and the corresponding fitting procedure for the city of Los Angeles are displayed in figure \ref{fig:ranking_2.png}.
Results for the estimation of entropy $h$ for the sampled cities can also be seen on a horizontal axis in figure \ref{fig:ranking_2.png}, showing how this measure introduces a clear sorting among our data.

The similarity networks were constructed fixing the threshold value to $0.018$, which corresponds to the 90\% confidence interval of the extrapolated values of $h$.
We chose to implement the clustering analysis in increasing subsets of our pool of cities, starting within a same region.
This way, it was easier to extract and visualise potential patterns or clusters of cities
sharing similar entropy levels.

\begin{figure}[h!]
\begin{center}
\includegraphics[width=0.50\textwidth]{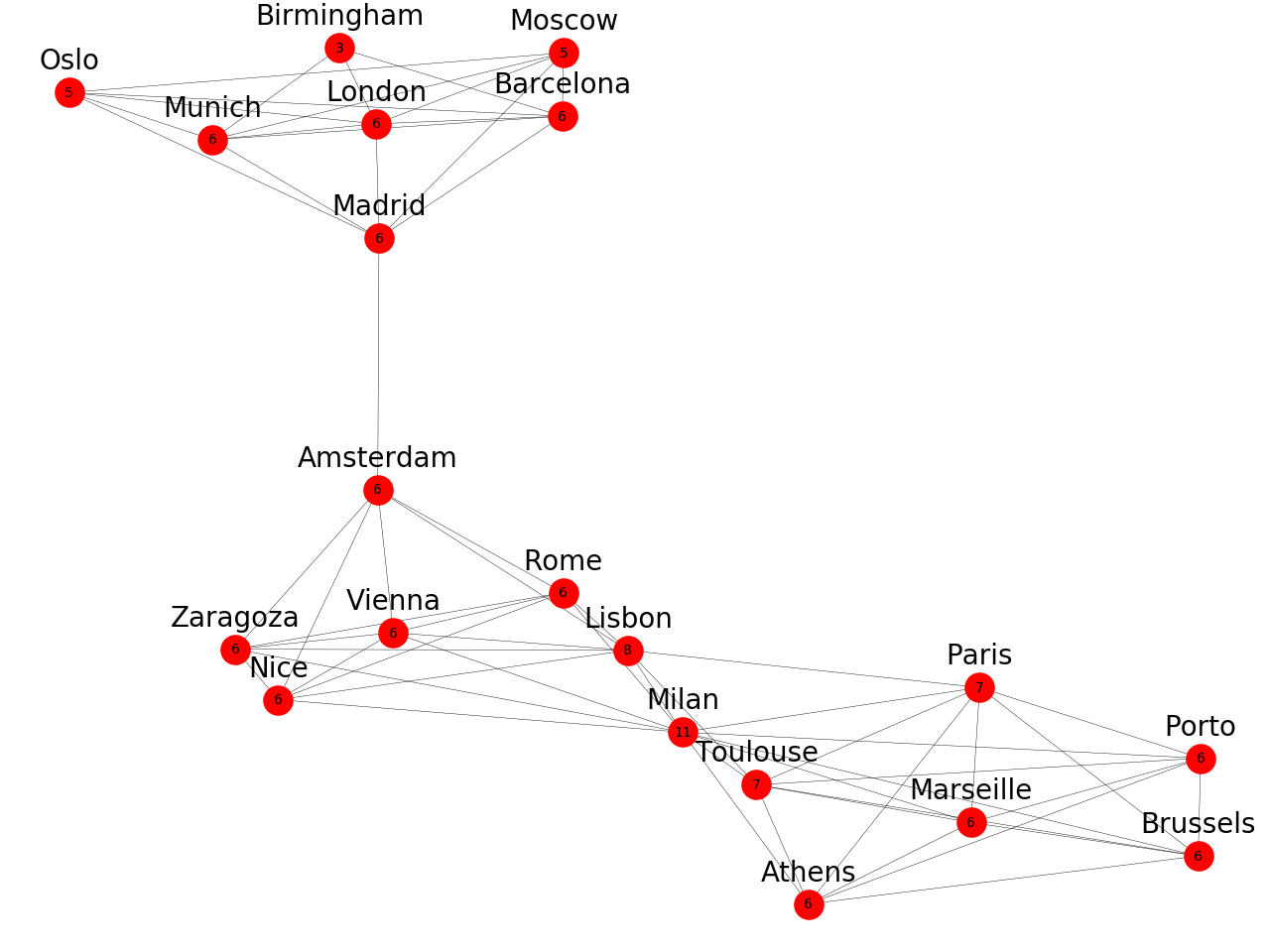}
\includegraphics[width=0.40\textwidth]{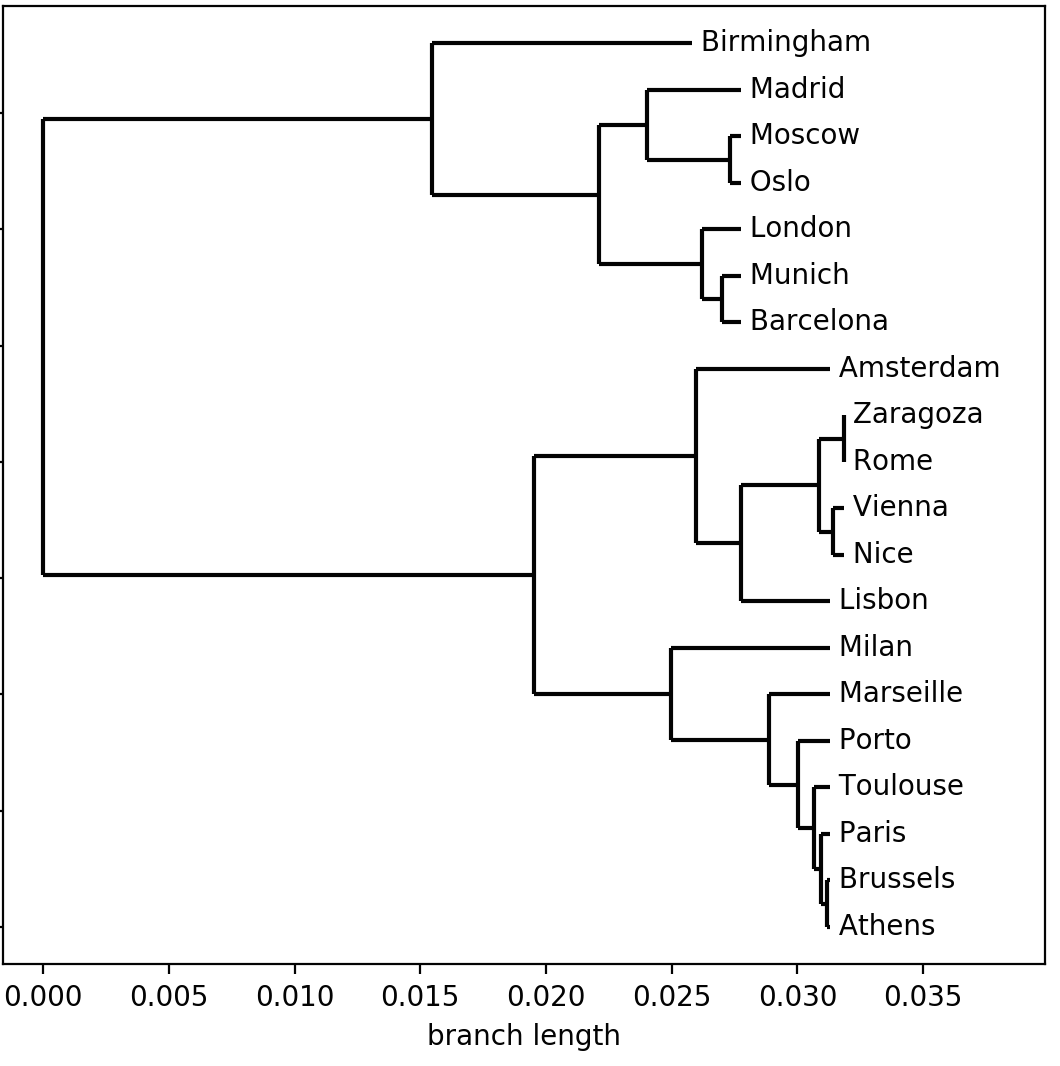}
\end{center}
\caption{\small 
\textbf{Left:} Proximity network of the considered European cities based on the value of $h$.
The edge lengths are not proportional to the levels of proximity between entropy values. 
\textbf{Right:}  %Phylogenetic-like tree
Dendrogram constructed using the UPGMA method applied to the distance matrix obtained in terms of the values of $h$. %The relative positions between pairs of cities is the most important information displayed by the tree.
The important information displayed by the tree is its topology. %In fact, the branch lengths can not be directly associated to the time since separation.
The tree features essentially two main morphological groups
corresponding to cities in Northern and Southern Europe.
%with an isolated group (Amsterdam and Vienna)
%corresponding to the cities at the edge between the two groups.
%Notice that the tree is unrooted, i.e., it does not require any hypothesis about common ancestors. What is important is the relative positions between pairs of languages. The branch lengths do not correspond to the actual distances in the distance matrix.
%upgma: unweighted pair group method with arithmetic mean
}
\label{fig:treesEu}
\end{figure}

We started by looking into European cities (figure \ref{fig:treesEu}).
Selected cities in Europe cluster in two main groups in the proximity network and %\textcolor{blue}{
corresponding dendrogram. 
The first one includes predominantly cities in Northern Europe, along with Barcelona and Madrid, which present lower levels of entropy. % Barcelona is a particularly interesting case, due to the extensive planning of Cerdá's orthogonal grid in the Nineteenth century, placing it along lower entropy cities. 
The second one includes mostly cities in Southern Europe, with higher entropy levels.
%}
% [1] The first one refers \textcolor{green}{predominantly to cities in Southern Europe, whereas the second includes cities in Northern Europe with Barcelona and Madrid, which present} lower levels of entropy.
%\textcolor{blue}{
The clustering displayed by the proximity network shows how Madrid and Amsterdam lie at the connection between both communities.
%}
% [2] The hierarchical clustering displayed by the dendrogram  shows similar characteristics \textcolor{green}{with an interesting branching in the group of the cities of the Southern Europe.}
%with interesting features, like the isolated group corresponding to the cities of Amsterdam and Vienna in between branches in Southern European cities, while lying at the edge between the two principal communities, along with Nice, in the network analysis.

\begin{figure}[h!]
\begin{center}
\includegraphics[width=0.59\textwidth]{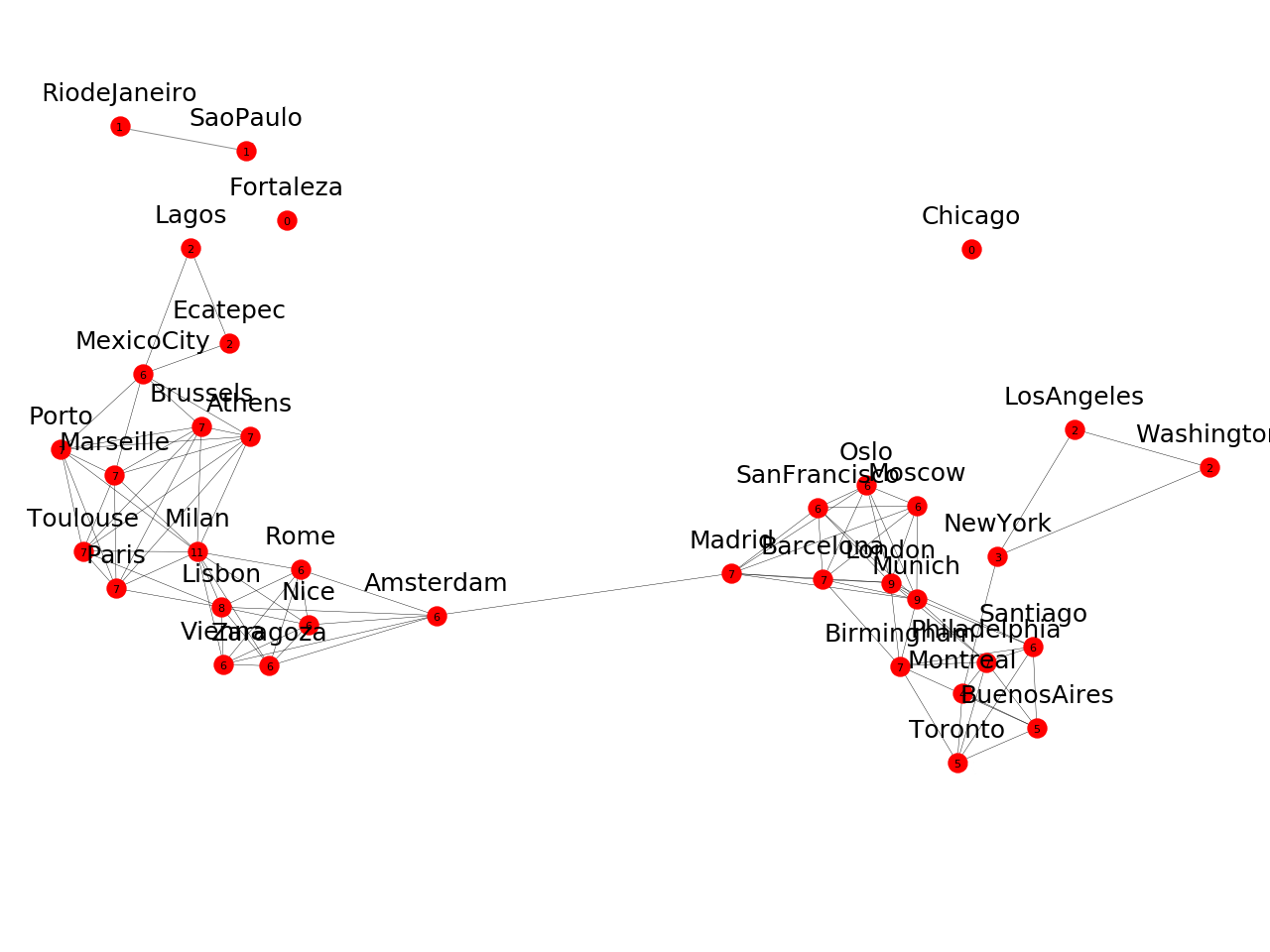}
\includegraphics[width=0.40\textwidth]{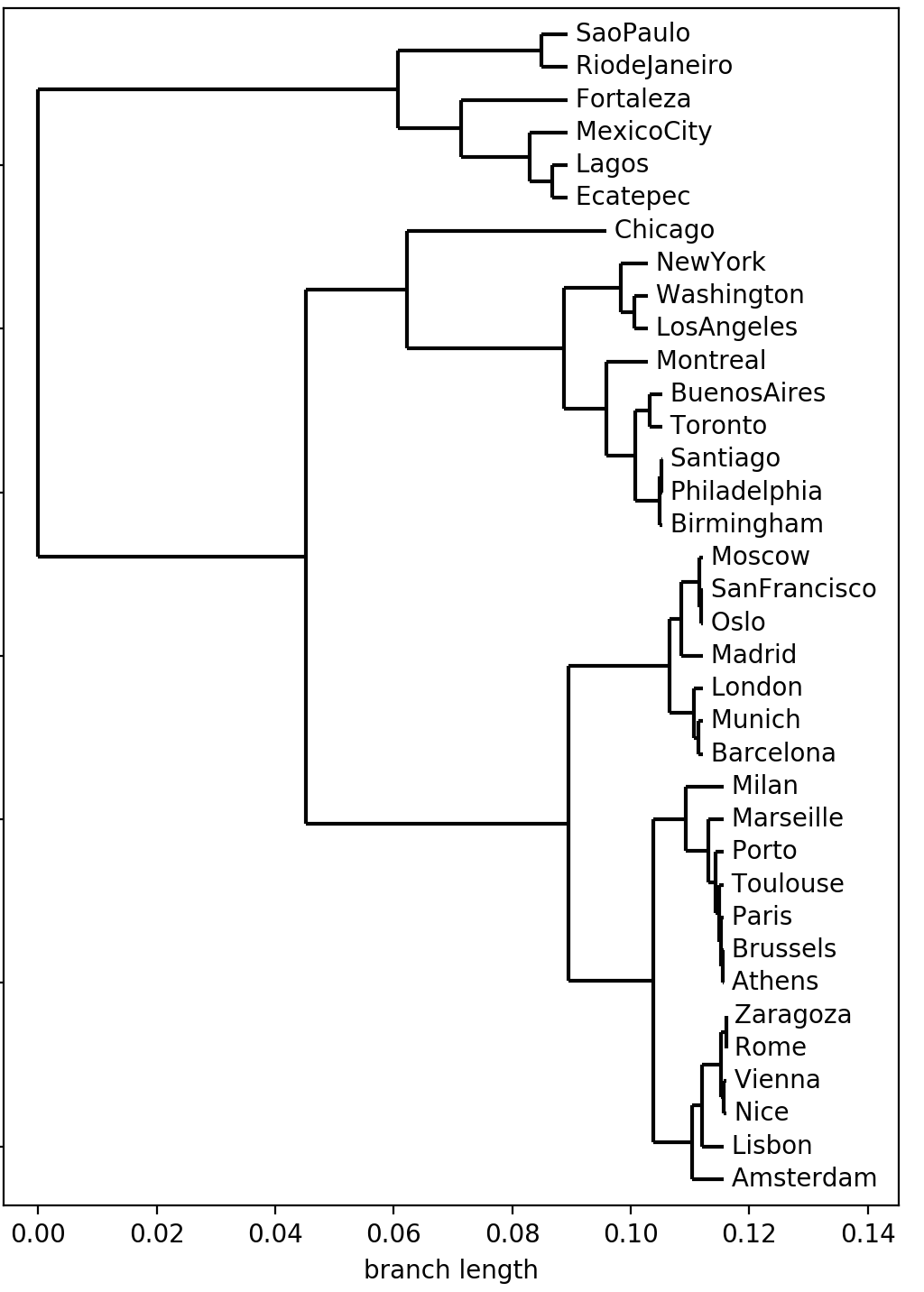}
\end{center}
\caption{\small Proximity network and dendrogram of the analysed European, North and South American cities, and Lagos in Africa.
}
\label{fig:tresEuAm}
\end{figure}

Next, we analysed the cities of Europe and the Americas along with Lagos in Africa (figure \ref{fig:tresEuAm}). 
We can distinguish 
different clusters in the proximity network.
While Brazilian cities São Paulo, Rio de Janeiro and Fortaleza remain as isolated clusters, other Latin American cities Mexico City, Ecatepec and Lagos in Africa form a small cluster joint to a large cluster dominated by cities of Southern Europe.
Another major cluster aggregates cities of Northern Europe and Canada, along with US city San Francisco, Spanish cities Barcelona and Madrid, and South American cities Buenos Aires and Santiago de Chile. 
A smaller connected cluster is formed by major cities in the United States, whereas Chicago stands as an outlier. %}
%\textcolor{blue}{
The dendrogram further clarifies these relations: Mexico City and Ecatepec along with Lagos share a common branch with the cluster comprised of Brazilian cities. 
Major US cities Chicago, Los Angeles, New York and Washington are placed in related branches, close to other North American cities (except San Francisco). Birmingham, Santiago and Buenos Aires relate to a same branch, as cities with the lowest entropy levels in their respective regions.
There is also a branch relating Northern European cities, Madrid and Barcelona, and a major cluster dominated by Southern European cities.
%}
%In the proximity network, we can distinguish \textcolor{green}{five different clusters. The first one is formed by cities selected in United States. The second one aggregates to the cities of Northern Europe cities from Canada, San Francisco, Buenos Aires e Santiago.  Finally we have a smaller cluster comprised of the Brazilian cities of Rio de Janeiro and São Paulo. 
%Most Northern European cities cluster with Barcelona and Madrid, along with San Francisco, a less gridded US city due to topographical constraints.
 %both heavily planned around orthogonal grids during Spanish colonization.
%The first one is formed by cities selected in Northern Europe and North America (US and Canada).
%A second cluster joints cities in Southern Europe, the South American cities of Buenos Aires and Santiago, and San Francisco, in North America.
%We have a smaller cluster comprised of Brazilian cities (Rio de Janeiro, São Paulo and Fortaleza) along with the African city of Lagos. 
%Finally, we have a very small cluster with two cities in Central America (Mexico City and Ecatepec).
%These results are transposed into the analysis of the corresponding tree, which, furthermore, shows some interesting details. \textcolor{green}{The two clusters of Mexican and Brazilian cities along with Lagos in Africa manifest similarity. Likewise, the cities of North America, with the exception of San Francisco, belong to the same branch.}

\begin{figure}[h!]
\begin{center}
\includegraphics[width=0.58\textwidth]{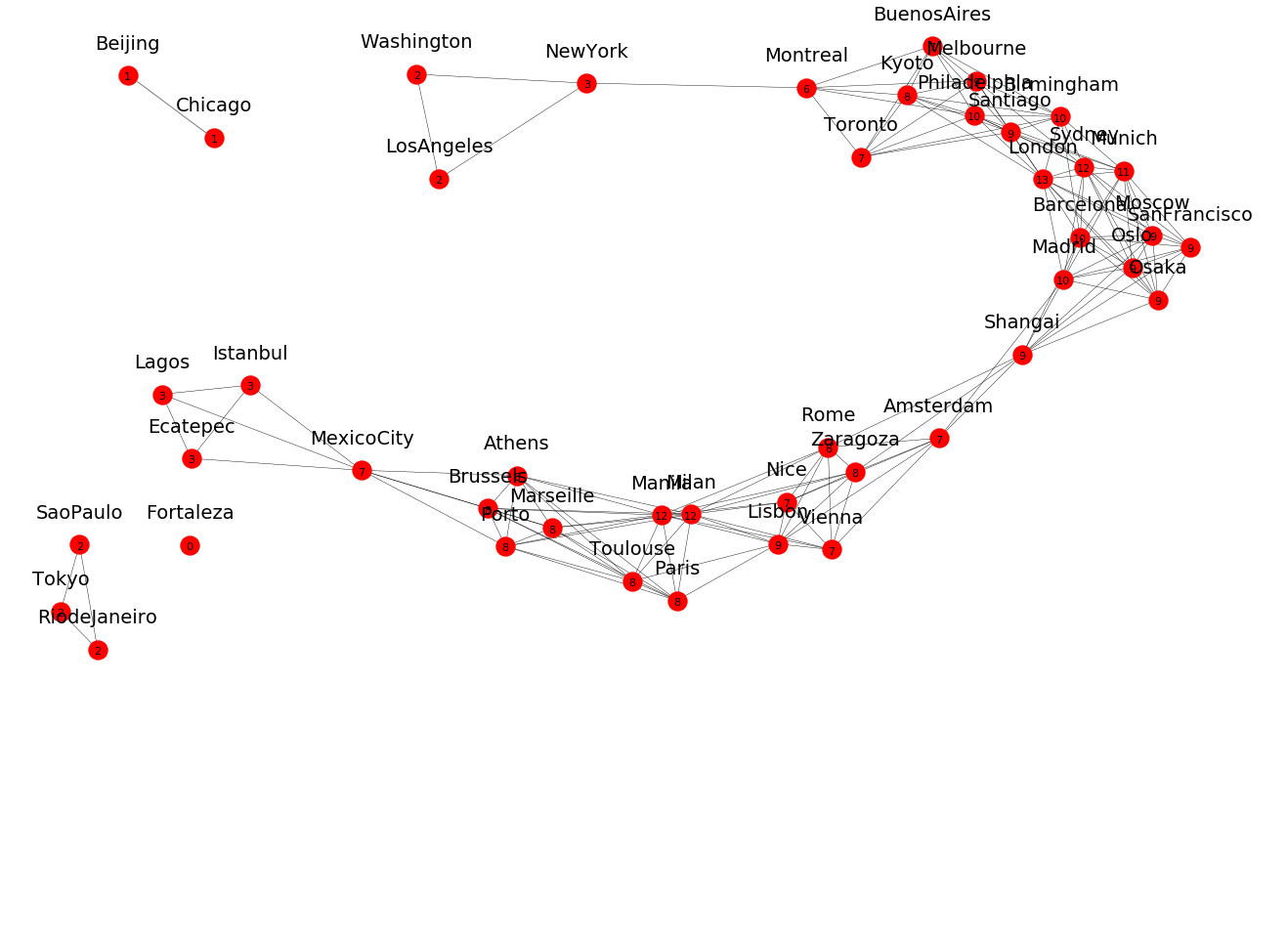}
\includegraphics[width=0.40\textwidth]{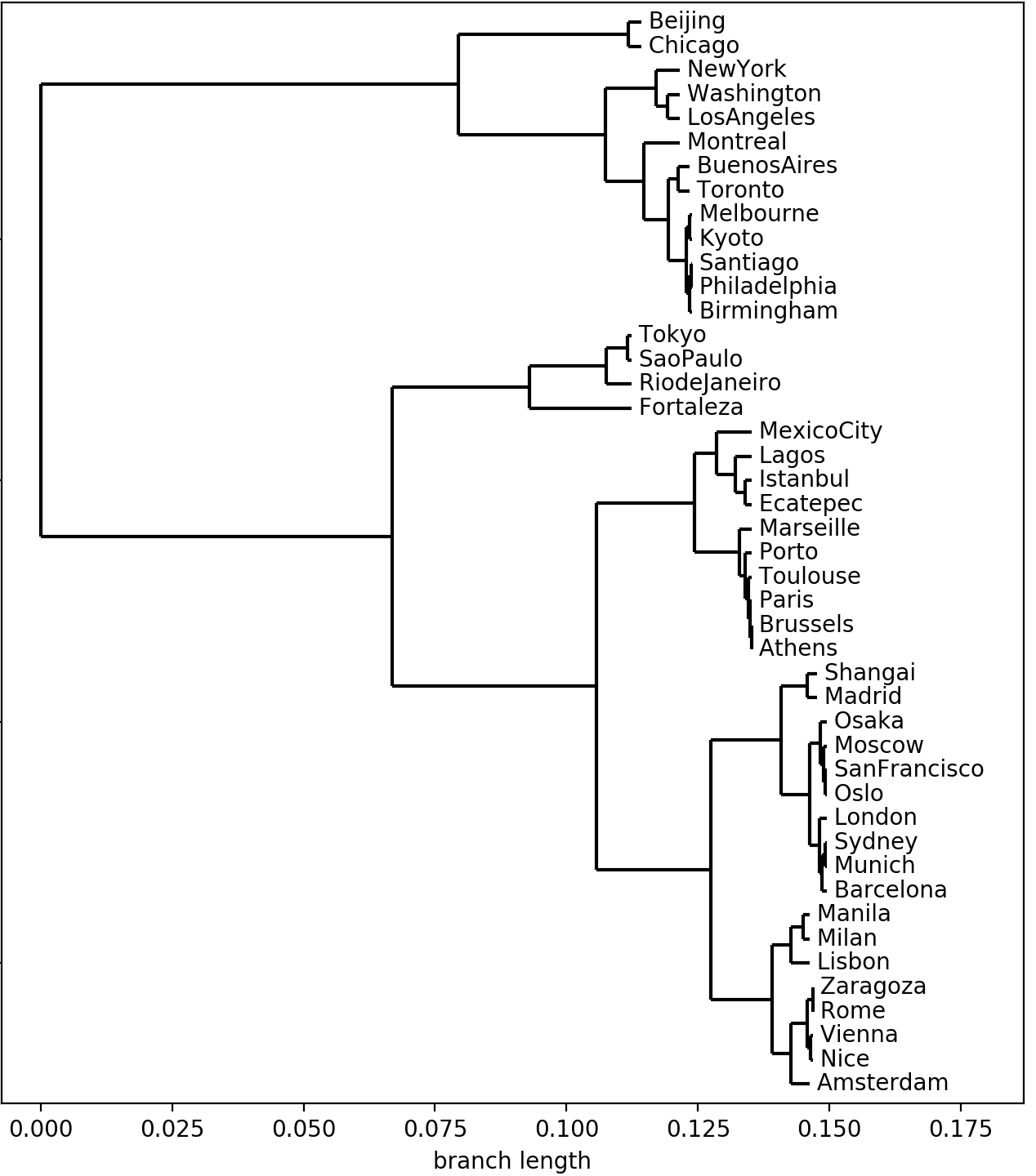}
\end{center}
\caption{\small Proximity network and dendrogram 
of North American, European, Asian, Oceanian, African and South American cities under analysis.}
\label{fig:tresEuAmAS}
\end{figure}

The concluding analysis joins together all the considered cities, adding the Asian and Oceanian data.
The number of clusters in the proximity network is similar to the previous analysis, %\textcolor{blue}{
with the addition of a new one with the most ordered cities, Beijing and Chicago.
%}
Apart from this fact, the community structure seems unchanged. Other Asian cities distribute themselves among pre-existing clusters: 
%With the exception of Beijing, which displays a very low level of entropy, 
most Asian cities join either the cluster dominated by Southern European cities or the cluster with most Latin American cities.
%\textcolor{blue}{
Furthermore, the network shows an interesting connectivity, from the most ordered cities Beijing and Chicago to major North American cities, then to a mixed cluster formed by cities from different regions sharing relatively low entropy levels, connected through Shanghai to a large cluster dominated by Southern European cities. This cluster in turn connects to the highest entropy groups, from Mexico City to Tokyo and Brazilian cities.
%}
The complete dendrogram can be seen in figure \ref{fig:tresEuAmAS}.

\section{Discussion}

What can the proximity networks and hierarchical clusters based on entropy values tell us about the \textit{cultural hypothesis?}
That means examining
the possibility that similarities in the ways of ordering built form can be explained either by (i) regional proximity, as cities from a geographically defined culture and identity (e.g. ‘American cities', ‘Italian cities', ‘Islamic cities');
or by (ii) similarities in the form of producing patterns historically shaped by %collective action and 
tradition in self-organised, bottom-up processes or by top-down agencies of self-regulation, allowing us to find elements in common even between different regions.
Of course, our analysis brings no value judgement in the sense of pointing out a certain level of entropy as desirable.
We may start by interpreting these differences in the light of the \textit{‘planned versus unplanned' dichotomy} so persistent in the urban imagination \cite{kostof1991city}. % (Kostof, 1991).

Beginning with the analysis of European cities, we found a subtle difference in levels of entropy between Northern and Southern cities. 
%Northern European cities were shown to have higher levels of order and predictability in their cellular arrangements than Southern cities.}
Northern European (i.e. Anglo-Saxon, Germanic and Russian) cities in our sample displayed in general lower levels of entropy -- from Birmingham (0.209) and Munich (0.225) to Amsterdam (0.254) and Vienna (0.263) -- than Southern (i.e. Latin European) cities, from Rome (0.260) to Paris (0.286) and Marseille (0.292), with the exception of Spanish cities Barcelona (0.227) and Madrid (0.240). 
While the analysed area of Madrid is composed as a patchwork, its parts are mostly regular in themselves. In turn, a large part of contemporary Barcelona was notoriously built according to Ildefonso Cerdà's 1859 \textit{Eixample} orthogonal plan. These features echo the Spanish tradition of regular grids deployed in colonized regions in Latin America, coupled with a strict alignment of buildings frontal facades, and contribute to set them apart from other Southern European cities.
Potential common traces in Northern cities include grids usually composed like a patchwork of partially regular areas (e.g. London, Munich, Amsterdam). 
This development pattern is frequently related to prior rural ownership and property boundaries. Regular grid sections relate to resources like land survey and delimitation based on measurement, prior to subdivision into building plots \cite{kostof1991city}. % (Kostof, 1991).
These cities also display considerable consistency in the building type adopted, leading to regularity in urban block surfaces.
Even though Munich’s historical core shows curved urban blocks, frontal and back facades are predominantly aligned. Geometric variation in the position of rear facades may be intense, combined with frequently sinuous urban blocks (e.g. Moscow, Vienna, Brussels).
%\textbf{On the one hand, Northern European cities like London or Amsterdam may have stronger rules guiding the local arrangements of buildings planning rules such as facade alignments} 
%–- even though they may vary widely in the geometry of urban blocks, due to historical shifts in emphases on order since the spontaneous urban formations of the Middle Ages to modern planning established mainly since the Nineteenth century.
% In Southern European cities, a number of pre-urban features have shaped the somewhat irregular form of urban blocks and street networks in historic cores, \textcolor{blue}{including medieval transformations of previously gridded Roman colonia settlements,} while asserting regularity in the form and position of buildings.
%MOST CITY ARE OF ROMAN ORIGIN, which means a very regular shape (from the 'castrum') which is hidden by the medieval interventions.

In turn, frequent curves in streets and block systems may follow medieval footpaths of previous open fields and rural field divisions related to landscape features (e.g. historic cores of Milan, Lisbon and Athens).
Practical modes of plot division and  building seem to closely relate to topography (e.g. Lisbon) and watercourses (e.g. Nice, Toulouse and Zaragoza).  
In these areas, buildings can be frequently strung along topographic lines and watercourses.
Despite such irregular features, there is considerable consistency in the position of frontal facades aligned along streets and open spaces.

To be sure, \textit{bottom-up processes of cellular aggregation} take morphogenetic paths involving randomness \cite{hillier1984social, batty2013}, trial and error \cite{alexander1964notes, alexander2003new}, and path dependence \cite{netto2017fabric}. %[on the concept of path dependence, see \cite{arthur1994increasing, arthur2013complexity}].
%(see Alexander, 1964; 2003; cf. Arthur, 1994; 2014).
These are processes where location decisions may influence the direction of subsequent decisions.
If an urban system shows positive feedback from a particular configuration, an increasing proportion of that choice increases the probability of another building being added in a similar way to the system, favouring the dominant pattern.
%Each decision in favour of a particular pattern increases the probability that the next selection will favour the dominant pattern (Arthur, 2014). 
This means that the built form system can phase-lock in a specific, path-dependent configuration.
Geometric consistencies resulting from trial and error processes and urban advantages triggered by increasing densities and decreasing internal distances \cite{netto2017social} % (Netto, 2017) 
can be reproduced as traditional modes of building. 
This process may eventually lead to institutionalised rules, like those prescribing particular building types, facade alignments or uniform setbacks even along originally unplanned street networks.

%may have to do with the passage from rural to urban landscapes, and how pre-urban land division may have shaped the irregular form of cities especially in the south.
%different methods of land division, fixing boundaries in relation to natural features, and the role of landscape features, or measuring with proper instruments to establish orthogonal relationships.
%This could suggest both regional affinities and common origins, local cross-fertilisation.

When we take the 45 cities into account,
we notice three main branches in the hierarchical clusters (figure \ref{fig:tresEuAmAS}, at a branch length around 0.075). 
A first cluster clearly emerges with cities with the lowest levels of entropy in the sample lower entropy cases. 
It is further divided into three initial branches. 
Beijing ($h=$0.111) and Chicago (0.116) 
have the lowest levels of entropy, and are in a branch of their own. 
Beijing is an exception in the Asian context, which generally has higher entropy values, 
from Shanghai (0.243) and Kyoto (0.206) to Tokyo (0.380).
Beijing is probably the most strictly planned city in China. Planning was implemented rigorously along cardinal directions (East, West, South, North) following a tradition traced back to early Ming dynasty (1368-1644 AD), in turn based on ‘regulations of construction’ from the Fifth century BC, as expressions of both regal power and social order \cite{wainwright2016}. Buildings and urban blocks frequently display regular forms and aligned facades. %despite variations in relative positions.
In turn, Chicago epitomises the US tradition of planning cities based on orthogonal grids – and it does so with great regularity in urban blocks and building surfaces.
%\textcolor{blue}{Northern European cities are likely to display similar levels of entropy -- from Birgmingham (0.209) and Munich (0.225) to Amsterdam (0.254) and Vienna (0.263).}

Other branches bifurcate into a group with major American cities New York (0.174), Washington (0.167) and Los Angeles (0.162), and configurations with the lowest entropy levels from other world regions, like Kyoto, Melbourne and Birmingham, along with other US/Canadian cities 
Montreal (0.190), Toronto (0.202) and Philadelphia (0.208).
Interestingly, Buenos Aires (0.198) and Santiago (0.209) cluster here, quite apart from other cities in Latin America, with high entropy levels.
This somewhat surprising result \textit{runs counter the first aspect of the cultural hypothesis:} the similarity in entropy levels for cities within a same culture or region.
This might have to do with the evolution of these cities in comparison to others in the Latin American region. 
Cities founded in the Sixteenth century by Spanish colonizers in the Americas were often created in a rigid orthogonal pattern, following the 1573 {\it Ordenanzas de Poblaciones}, % which can be considered as 
the first code of urbanism of the early modern period in the West. This was the case for Santiago and particularly Buenos Aires, with its plain topography %and later foundation
\cite{salcedo1996urbanismo}.
These areas became the historical and economic core of these cities, with high density and compact patterns of built form. As these cities expanded, patchworks were added around the core's regular structure, adding entropy to the mix. Nevertheless, the levels of order in those central configurations are felt in the analysis, %detaching these cities from high entropy morphologies found in other Latin American cities, 
bringing them to closer to cities with higher levels of order in built form, like Toronto and Philadelphia.

The second cluster %at 0.075 branch length 
highlights the highest entropy group in the sample, comprised of Brazilian cities Rio de Janeiro and São Paulo, in Latin America ($h=$ 0.391 and 0.382, respectively), followed closely by  Tokyo (0.380) and another Brazilian city, Fortaleza (0.347).
%, as clarified by the linear entropy scale (Figure 6).

The third major cluster divides into communities from high to middle entropy. This cluster further bifurcates into a group with slightly lower entropy levels, Mexico City (0.303) and Ecatepec (0.320) in Mexico, Istanbul (0.322) in Turkey, and Lagos in Africa (0.315).
%and Manila in the Philippines (0.390). 
Another bifurcating branch between the opposing clusters is comprised of Southern European cities Marseilles, Porto, Toulouse, Athens and Paris, along with Brussels. 
%, a part of the French community of Belgium, separate from the Flemish region.
A final large branch of middle to lower entropy cities bifurcates into cities from diverse regions, like Manila (0.274) in the Philippines, Milan (0.277), Rome (0.260), Lisbon (0.268), Zaragoza (0.260) and Nice (0.262) in Southern Europe, and Amsterdam and Vienna in Northern Europe; and into more diverse groups with lower entropy cities, like Shanghai %(0.243) 
and Madrid%(0.240)
, Moscow (0.231) and San Francisco (0.233), Sydney (0.224) and Munich, London (0.223) and Barcelona. %(0.229).

These distinct clusters show that \textit{we cannot associate particular levels of entropy exclusively with particular regions, a first possibility of verifying the cultural hypothesis}.
We have to ask ourselves what in different regions could have triggered similar entropy levels.
The idea of a planned-unplanned dichotomy suggests that we should look into the actual evolution and planning conditions existing (or not) in these different cities, many of them having faced considerable growth in the twentieth century. 
We checked the existence of modern planning rules that act specifically upon built form, namely:
(1) Land parcelling: how land is divided into urban plots, and whether there are rules guiding the shape and regularity of plots.
(2) The layout of urban blocks and streets: what are the rules for layouts – say, whether they impose orthogonal systems or ‘planned picturesque’ systems like intentionally curved and varied block shapes and street networks.
(3) Regulations on building design and location: whether there are rules that specify the position of buildings in plots (e.g. frontal and lateral setbacks), and in relation to neighbouring buildings.
We examined the legislation in emblematic cases in Turkey, Nigeria, China, Brazil, Mexico, United States, England and The Netherlands. 
We found something that goes counter the planned-unplanned account of ordered and disordered cities: 
cities which have top-down planning may also exhibit high built form entropy.
They do have rules and government agencies that regulate building and urbanisation. 

But how can high entropy in built form be somehow influenced by top-down rules?
We found that cities from different regions -- namely, Brazil, Nigeria, Mexico and Turkey -- may have certain aspects of planning in common, which allow great variation in built form to come into being. 
For instance, these cities share emphases on parcel-based, piecemeal developments. 
New urbanised areas are mostly exempt from requirements to keep connections to neighbouring areas, including street continuity and grid alignment.
Another crucial instance here is how individual buildings can be positioned in their plots. 
Some regulations may enforce frontal and lateral setbacks, and define rules like increasing setbacks as buildings grow taller. 
\textit{Simple local rules focused exclusively on individual buildings} rather than coordinated construction among nearest neighbours \textit{lead to a high level of fragmentation} in built form.

Going a step further, whole areas in these cities are urbanised and built by people's own hands in informal settlements, hence apart from planning regulations. 
This is especially the case throughout the Twentieth century, when cities in developing countries experienced fast growth.
We are likely to find high entropy mostly associated with variation in the shape of urban blocks (related to angular variation in surrounding streets) in those settlements. %That is also the case for compact urban blocks with continuous facades between neighbouring buildings. 
In short, parcel-based, piecemeal developments, patchworks of diverse blocks and street networks, and fragmented built form are key features of highly entropic urban landscapes.

\begin{figure}[h!]
\begin{center}
\includegraphics[width=0.8\textwidth]{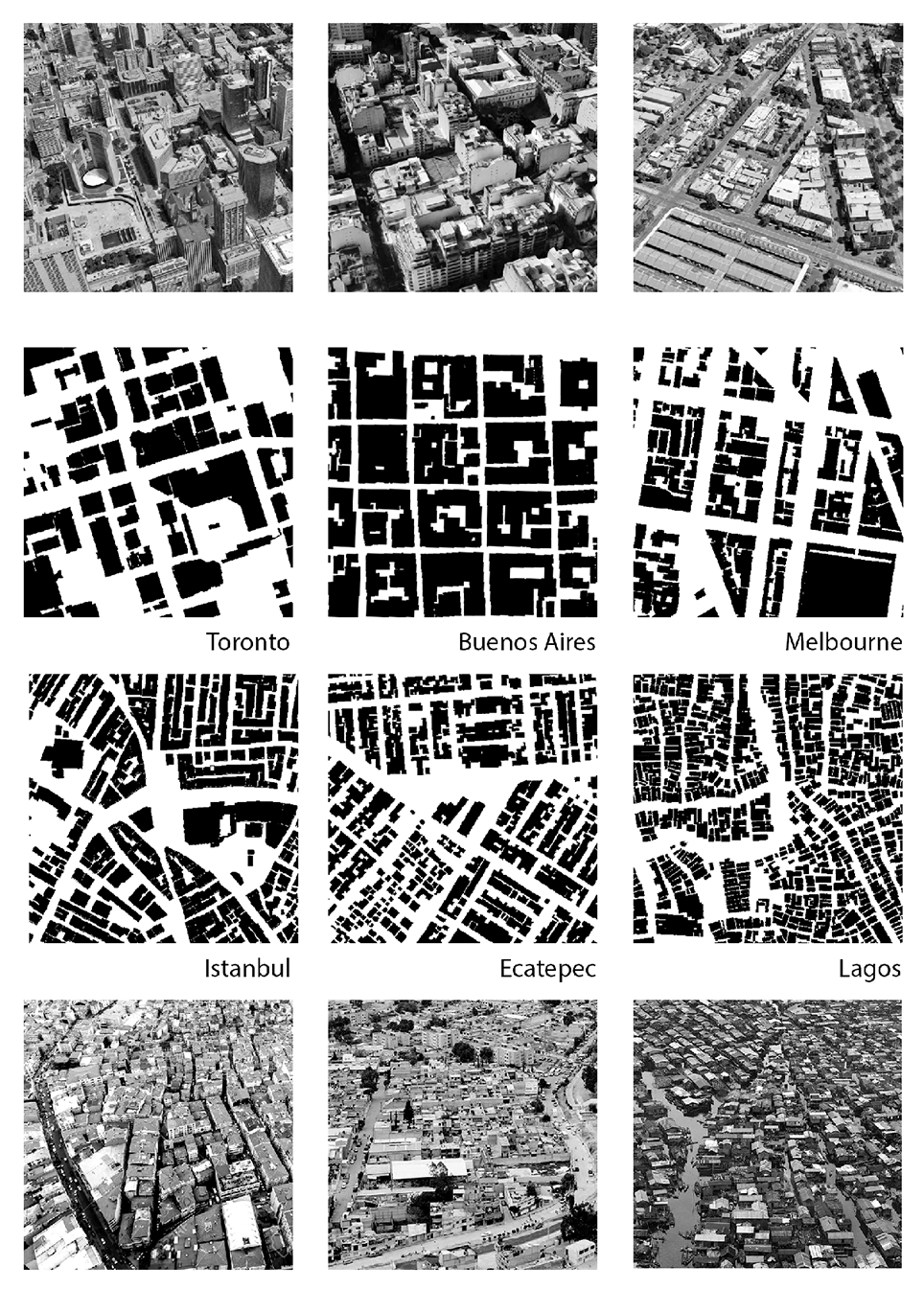}
\end{center}
\caption{\small Urban sections (500x500m) with similar entropy levels, different spatial configurations: Toronto ($h=$ 0.202), Buenos Aires (0.198) and Melbourne (0.207); and Istanbul ($h=$ 0.322), Ecatepec (0.320) and Lagos (0.315).}
\label{Figure_10.png}
\end{figure}

All this shows that cities from distinct regions may share similar entropy levels as far as built form is concerned. Their typical combinations of cells might be different, and they might share neither geographical proximity nor common historical roots, but they still can contain similar levels of disorder, as captured by our measure (figure 10).
This suggests certain common traits between different regional cultures  shaping how built form is ordered. 
%\textit{regions do not have entropy values necessarily different from other regions.}
%a single  regional explanation cannot respond to all cases.
%In addition, the relationship of built form and street networks may also help to shape cities into higher levels of entropy in bottom-up, ‘spontaneously’ emergent urban fabrics, like in the medieval cores of European cities found in Southern Europe, like French and Italian cities emblematic for urban structures in close relation with topographic features. 

That said, even though regions do not have entropy values necessarily different from others, \textit{individual regions do seem to converge around certain values.}
This interesting pattern emerges once we visually distribute a classification of the 45 cities 
according to increasing entropy values on a global map (figure \ref{fig:Fig3_new2.png}). 
%Northern cities in general display lower entropy levels.
Some regions show higher levels of regularity and predictability in built form systems than others. 
We suggest that our measure seems to capture spatial information potentially related to \textit{different emphases on order and coordination latent in different planning cultures}, the second aspect of the cultural hypothesis seen above.
%This suggests the possibility of a role for planning cultures shaping how built form is ordered.
%%%%% Also, deeper historical and local contingencies might be at play, and they must be carefully taken into account.

How do these findings on regional differences compare with previous studies, based on different spatial entities and methods? 
We have seen that Medeiros’s \cite{medeiros2013urbis} analysis of street networks %of 164 cities 
based on betweenness centrality and topological depth identified clusters of US/Canadian cities with the highest levels of accessibility, in contrast with Brazilian cities in South America, followed by European cities.
Largely echoing Medeiros's findings, %through a binary option of ‘planned according to clear organizing principles’ or evolving ‘organically through accretion’, 
Boeing \cite{boeing2019urban} explored angular orientation entropy 
and grid order indicators %in 100 street networks 
to identify % trends worldwide
US/Canadian cities with the lowest orientation entropy. 
%Outside the US, Kyoto appears more grid like also shows low orientation entropy. 
European cities also exhibit higher orientation entropy than Latin American cities. 
Louf and Barthelemy’s \cite{louf2014typology} %combined topology and geometry to generate a 
classification based on block areas 
%\textcolor{blue}{based on ratios of block areas as proxy to shape} 
%\textcolor{green}{based on blocks area ratio as proxy to shape} 
%extracted from street networks in 131 city centres, based on the ratio of block area and the area of the circumscribed circle, while unable to preserve the actual shape of blocks and take account of configurations of built form, and does not grasp information encoded in relations between urban blocks as a cellular approach can, the method 
also identifies differences between most American cities and European cities.

US/Canadian cities display low entropy in our analysis as well, 
but our results on European cities differ from those studies. 
Consistency in built form in European cities brings entropy to lower levels. Our approach was also able to identify differences between Northern and Southern European cities. 
%,possibly in connection with %urbanisation and original rural structures and the use of resources like land survey and measurement prior to plot division, as opposed to practical forms of division related to topography \cite{kostof1991city}. 
%may be felt particularly in Southern European cities. 
% and the remarkable %dominance of irregular shapes and sinuosity of urban blocks in historic cores.
In their turn, São Paulo and Rome exhibit the highest entropy levels in Boeing's study. 
In our approach, despite the varied shapes in its block system, Rome’s consistency around aligned buildings lowers its entropy to a level far from São Paulo. 
Interestingly, these different approaches converge about Brazilian cities: they exhibit the lowest average betweenness centrality, and highest orientation entropy and built form entropy in these different samples -- probably due to an extraordinary variation in the position of buildings in plots, coupled with fragmented grid patchworks.
Nevertheless, the differences between findings are clearly related to differences between the morphologies of street networks and built form systems: the fact that a same street network can support endlessly different configurations of buildings.
Levels of order in street networks 
do not necessary cause low entropy in built form.
%Levels of order and entropy in streets and built form are not necessarily correlated.

%Certain cultures may place stronger emphasis on rules for local coordination of building production and arrangement of buildings along urban blocks, either emerging from of agents and/or top down institutions with positions in coordinating the. 
%As a consequence of little spatial coordination between buildings in architectural and urban design, built form and urban landscapes in those countries find great variation and fragmentation.
%Therefore, as aforementioned, on the one hand the entropy measure applied to local cellular arrangements of built form does not necessarily grasp spatial cultures according to specific regions. Clearly, many different configuration microstates may correspond to the same value of Shannon entropy/information.
%Certain cultures may place stronger emphasis on rules for local coordination of building production and arrangement of buildings along urban blocks, either emerging from of agents and/or top down institutions with positions in coordinating the. 

\section{Conclusion}

In this paper,
we developed an approach to spatial information based on Shannon entropy.
The approach was designed to (1) measure the entropy characterising levels of order and disorder in cellular configurations present in 45 cities around the world.
(2) we applied the method to investigate the hypothesis of ‘spatial cultures' as ways of ordering urban form. 
Put another way, we verified whether the entropy measure could accurately grasp features and differences in built form systems; then we looked for traces of ‘information signatures’ potentially consistent with specific regions or cultures.
This method is intended as a step towards a more precise understanding of \textit{spatial cultures as emergent patterns} -- i.e. how typical configurations of built form emerge from local rules of aggregation active at the scale of cellular configurations.
%We argued that, due to its reach into levels of randomness and order in configurations and its natural relation to forms of encoding messages in both symbolic and material media, Shannon's entropy measure was particularly suited for grasping information in ways of ordering built form related to specific cultures.

Of course, any search for ‘information signatures' of spatial cultures embodied in the tangible spatiality of cities faces certain risks:
(a) Different cultures or regions may not have distinct ways of ordering space. In other words, there could be no ‘spatial cultures’ related to regions or even enough differences between cities to be associated with a particular culture. 
(b) Spatial cultures may well have specific information signatures, but these may not be encoded at the scale of local cellular configurations of built form systems.
(c) In case the possibilities above were wrong, a measure of spatial information based on Shannon's entropy in cellular configurations may not be precise enough to capture information signatures or even qualitative differences in configurations.

In the research process, our method allowed us to find the distribution and clustering of cities around certain values of built form entropy. 
%We noticed the possibility that results could shed light on a classic idea in urban studies, related to differences in the ways different regions and cultures shape their cities and order their spaces. 
%The measure was capable of identifying consistencies apparently similar, but not to \textit{regional} cultures. Clusters were closer to differences between \textit{planning} cultures.
We would like to conclude our work discussing such entropy values
and clusters in connection with characteristics of these cities, including %non-physical 
aspects %that compose their ‘culture', 
%like values in planning 
particularly %\textcolor{red}{
related to what we called ‘cultural hypothesis': 
%the possibility that different regions and cultures find different ways of ordering space according.
the idea that similarities in the ways of ordering built form can be explained by 
(i) regional proximity, as geographically defined cultures and identities, or by
(ii) common features in urban morphogenesis shared by distinct regions, say rules in similar ‘planning cultures'.
That meant looking for reasons of non-contingent similarities and differences between cities.

The usual association of bottom-up processes of spatial production in disordered, unplanned cities, as opposed to top-down processes of spatial production in ordered, planned cities, suggested that we should look specifically into planning rules guiding built form. 
We found that the ‘planned/unplanned' dichotomy in urban studies may have been valid in pre-modern periods of certain urban cultures, %(e.g. the historical formation of urban cores in Northern and Southern European cities), 
but it seems of limited explanatory power once we consider modern and contemporary planning. %in developing countries. 
%And we found something that goes counter the planned/unplanned account of ordered/disordered cities: 
Cities with top-down planning may also have high built form entropy. 
%\textcolor{red}{Cities with the higher levels of order and predictability in their cellular arrangements display considerable regularity in urban block surfaces, with a high consistency in the position of frontal facades aligned along streets.
%Cities with slightly lower built form entropy seem frequently composed like patchworks of partially regular areas (e.g. London, Berlin, Amsterdam), likely to be related to prior rural ownership and property boundaries, land survey and delimitation based on measurement, prior to subdivision into building plots. 
%In turn, cities with built form entropy around ‘medium values’ [VALUES] may share certain pre-urban conditions closely related to previous footpaths connecting systems of towns and settlements, and landscape features like topography (e.g. Lisbon) and watercourses (e.g. Nice, Toulouse, Zaragoza). 
%These features, coupled with more practical modes of plot division, may have led to irregular urban blocks especially in historic, dense urban cores, while allowing regularity in the shape and position of buildings strung along topographic lines and watercourses (e.g. historic cores of Milan, Lisbon, Toulouse and Athens).}

A key difference lies in the kind of rules applied and how they deal with buildings.
Cities with high entropy in different regions in our sample seem to have in common rules that focus mostly on individual buildings, allowing great variation in how they are placed in plots and blocks, including increasing lateral and frontal setbacks as buildings grow taller.
This focus may happen for specific periods of their histories -- long enough to shape the evolution and morphology of large portions of these cities. 
This is the case of planning in countries like Brazil, Mexico, Nigeria and Turkey, especially %during periods 
when the analysed cities faced fast growth in the Twentieth century. 
In short, we found that \textit{simple local rules centred on individual buildings rather than coordinated construction lead to high fragmentation in ensembles of built form}. 
These rules are frequently coupled with piecemeal developments and grid patchworks, including informal settlements, shaping a visible fragmentation of urban landscapes.
\\

Our analysis brings other findings. %\textcolor{blue}{
First, proximity networks and hierarchical clusters show similarities in cities from different regions 
(e.g. high entropy cities including São Paulo, Tokyo, Istanbul and Lagos), 
with close entropy values even if they have geometrically distinct arrangements (figure 10). %\ref{fig:Figure_10.png}).
%Clearly, many different configurations may correspond to the same value of Shannon's entropy.
%Lagos in Nigeria and Fortaleza in Brazil are very different spatially, but have very close entropy levels. 
This suggests that the measure does not necessarily generate specific values as exclusive ‘information signatures' for each region, a first possibility of verifying the cultural hypothesis.
%}
%\textcolor{red}{Even thought the measure found similar entropy levels in cities of a same region,} the fact that cities from different regions shared similar entropy levels mean that we could not use entropy to distinguish between regions.

Second, despite that fact, the measure seems to capture something of the ‘planning culture' of these regions. 
%\textcolor{red}{We mean here different emphases on order and rules guiding the construction of buildings and urban blocks in these cities. 
%For instance, the fact that many American cities were built based on orthogonal grids is well known \cite{Medeiros2013, boeing2019urban} (Kostof, 1991).}
%\textcolor{blue}{Cities with high entropy in built form, on the other hand, actually seem to have something in common: they do have top-down planning and rules, but %have more planning cultures, especially related to the absence of rules ordering blocks, streets, or the position of buildings in the lots –- say, like fixing the juxtaposition or alignment with neighbouring buildings. Such 
%rules focused on urban form as an ensemble are frequently missing from those planning traditions. Instead, rules are mostly focused on individual buildings and the problem of habitability.}
%\textit{What these cities seem to share is similar levels of entropy in their own ways of ordering built form}, which allow us to group them together despite regional and societal differences.
We found higher frequencies of certain regular arrangements %usually triggered by regularity  
in cities with top-down planning coupled with a strong focus on rules for coordinated modular construction, each building adjusting and aligning to those around,
%levels of coordination between individual buildings design solutions 
taking into account systemic consequences of ensembles of built form.
The high frequency of certain arrangements can also be found in cases of bottom-up processes of cellular aggregation potentially involving path dependence, i.e. built form systems locked into specific configurations, reproduced in traditional modes of building -- patterns that can be eventually institutionalised into formal planning rules. 
This seems to be the case especially in the passage from pre-modern to modern urbanisation of European cities.
On the other hand, we found plenty of variation in cellular aggregations in urban cultures %less fixated on rationalised, rectilinear planning, 
that allow the construction of buildings in uncoordinated actions between individual developers.
This clearly leads to less regularity and higher unpredictability in what surrounding built forms will be like as cities grow.
Summing up, in both top-down and bottom-up form-making processes,
local rules %\textcolor{blue}{engendered either by collective action in tradition and path dependence, or coordinated by institutionalised agencies and rules, or both,}
guiding how to position buildings in relation to others seem to trigger bifurcated developments as the built form system evolves in size and complexity, %(Batty, 2013:246), 
leading either into greater consistency or into greater fragmentation.
But that is not the whole story, of course. We may find many possibilities in between those archetypal paths, 
or combinations of them %in cities where both trends can be found, either present 
in different parts of cities,
like patchworks, 
or intermingled in layers of ordered and disordered aggregations -- say, the iconic case of Manhattan, based on the regularity of a gridiron street layout, and planning rules that made room for enormous variation in built form.

Third, although regions do not necessarily have exclusive values of built form entropy, \textit{individual regions do seem to converge around certain values.}
Our results show certain consistencies, grouping cities from a same region (e.g. Brazilian cities, American cities). To use Hillier's words \cite{hillier1989architecture}, this echoes the idea that societies %order their spatial milieu in order to 
create their own spatial cultures -- their distinctive ways of ordering space and shaping cities. 
%the global distribution of entropy values shows a certain pattern (figure \ref{fig:Fig3_new2.png}), with cities in the global south generally having higher entropy values.}
Such finding needs to be further examined through a larger sample of cities and comparisons with other approaches, along the lines we explored above.
Of course, deep historical conditions and local contingencies are likely be at play, and must be carefully taken into account. 

Finally, differences between results obtained from street network-based measures %like between betweeenness centrality and orientation entropy and 
and our measure of entropy shed light on the \textit{potential dissociation between the morphology of streets and the morphology of built form systems} in every city: the endless %possibilities there is between the morphology of streets and the 
combinatorial possibilities of configurations of buildings, missing from street network approaches, add complexity to urban phenomena and suggest the need for a renewed interest in built form systems.
%}
\\

%\textcolor{green}{%NAO SE SE VALE A PENA:
Our sample is not a random set, which would be impossible due to the lack of information on building footprints in many cities and countries.
%therefore it does not represent a broad cross-section of different regions, cultures, and development paradigms.
%Of course, a larger sample of cities coupled with the examination of their planning regulations are needed for more conclusive results. 
%Our approach can also be geared to identify combinatorial arrangements typical of particular regional or planning cultures, which will be the subject of a following work.
%%%  E D G A R D O : THE FOLLOWING SENTENCE WAS ALREADY IN OUR COSIT PAPER:
%\textcolor{green}{
Methodologically, at the present stage, our approach takes account of the spatial information latent in the arrangements of cells capturing relations of proximity, but eventually missing some correlations at large distances.
%}
On the one hand, %our method theoretically captures all the 
%different scales, but without hierarchy between them.
cellular growth shapes larger structures as fundamental features of cities -- a subject explored in other works \cite{haken2003face, hillier2012genetic, battyetal1989, batty2013}. 
On the other hand, humans have a clear hierarchical reading \cite{alexander1964city}, 
and structures at larger scales seem to have more weight than structures at smaller scales to differentiate objects.
Further development of this research will look into broader spatial structures in cities by introducing %more sophisticated 
measures of statistical complexity.
In addition, we wish to expand this approach to other forms of physical information, such as three-dimensional differences between buildings, physical cues and landmarks \cite{couclelisetal1987, stern1999environmental}. 

Even though there is no value judgement in our work or claims of particular levels of entropy as desirable, different levels of built form entropy may well trigger different cognitive and practical responses from people. Higher degrees of entropy may be associated with spatial and visual surprises in navigation. Surprises can be considered desirable by some, as famously suggested by Camillo Sitte \cite{sitte1979art}
and explored by Gordon Cullen's \cite{cullen1995concise} concept of ‘serial vision'.
Notwithstanding, empirical studies in spatial cognition and neuroscience have shown that certain regularities and alignment effects (say, between paths or objects like buildings, or triggered by cardinal directions) improve our judgement of relative direction in navigation and our capacity to determine the position of objects in a surrounding area, affecting intelligibility and our memory of the built environment \cite{mcnamara2003egocentric, frankenstein2012map, ekstrom2018human}.
%(McNamara et al., 2003; Frankenstein et al., 2012; Ekstrom et al., 2018).
Effects of urban form on cognition are a hot research topic and could benefit from explorations into entropy and regularity in physical space as informational features in navigation \cite{haken2003face, haken2014information, hillier1996space}. 
Furthermore, human knowledge of spatial properties and patterns goes beyond physical information and can integrate configurational, visual and semantic aspects of an urban environment \cite{ mulligann2011analyzing, haken2014information, netto2018cities}. 
More work is needed to understand how physical information %, namely in the form of semantic information, 
is associated with non-physical information, and is enacted by social agents making decisions and cooperating in cities.

\section{Acknowledgements}
%This work... \\
%Website: \url{http://www.sunrise-setting.co.uk}
We would like to thank the following researchers, architects and urban designers for their support in the analysis of planning regulations in a number of cities around the world: Cynthia Adeokun (Lagos and London), Ilgi Toprak (Istanbul), Mayra Gamboa Gonzáles, Juan Ángel Demerutis Arenas and Claudia Ortiz-Chao (Mexico City and Ecatepec), Tatiana Rivera Pab\'on (Buenos Aires and Santiago),
Akkie Van Ness (Oslo and Amsterdam) and Chaogui Kang (Chinese cities). We also thank Lilian Laranja  
for discussions on culture and built form entropy. 
Any errors in interpretation are the authors' responsibility. 
%Imprecision in interpretation is our own.

%\begin{thebibliography}{99}
%\bibitem[Kopka and Daly(2003)]{R1}
%Kopka~H and Daly~PW (2003) \textit{A Guide to \LaTeX}, 4th~edn.
%Addison-Wesley.
%\bibitem[Lamport(1994)]{R2}
%Lamport~L (1994) \textit{\LaTeX: a Document Preparation System},2nd~edn. Addison-Wesley.
%\bibitem[Mittelbach and Goossens(2004)]{R3}
%Mittelbach~F and Goossens~M (2004) \textit{The \LaTeX\ Companion},
%2nd~edn. Addison-Wesley.
%\end{thebibliography}

% ADD BIBLIOGRAPHY
% Conzen MRG (1960) Alnwick, Northumberland: a study in town-plan analysis. Transactions and Papers, Institute of British Geographers, no 27

\bibliographystyle{plainurl}
\bibliography{refs_01.bib}
%\bibliography{refs_01}

\end{document}